# Digital Twinning of the Human Ventricular Activation Sequence to Clinical 12-lead ECGs and Magnetic Resonance Imaging Using Realistic Purkinje Networks for in Silico Clinical Trials

Julia Camps[1,*,a], Lucas Arantes Berg[1,a], Zhinuo Jenny Wang[1], Rafael Sebastian[3], Leto Luana Riebel[1], Ruben Doste[1], Xin Zhou[1], Rafael Sachetto[4], James Coleman[1], Brodie Lawson[5], Vicente Grau[1], Kevin Burrage[1,5], Alfonso Bueno-Orovio[1], Rodrigo Weber[2], Blanca Rodriguez[1]

[1]University of Oxford, Oxford, United Kingdom

[2]Federal University of Juiz de Fora, Juiz de Fora, MG, Brazil

[3]University of Valencia, Valencia, Spain

[4]Universidade Federal de São João del Rei, São João del Rei, MG, Brazil

[5]Queensland University of Technology, Brisbane, Australia

* Corresponding authors:
Julia Camps
julia.camps@cs.ox.ac.uk

[a] Julia Camps and Lucas Arantes Berg made equal contributions to this work.

**Cardiac digital twin; Purkinje network; Bayesian inference; Eikonal model; Monodomain model; Cardiac Magnetic Resonance; Electrocardiogram**

## Abstract

Cardiac in silico clinical trials can virtually assess the safety and efficacy of therapies using human-based modelling and simulation. These technologies can provide mechanistic explanations for clinically observed pathological behaviour. Designing virtual cohorts for in silico trials requires exploiting clinical data to capture the physiological variability in the human population. The clinical characterisation of ventricular activation and the Purkinje network is challenging, especially non-invasively. Our study aims to present a novel digital twinning pipeline that can efficiently generate and integrate Purkinje networks into human multiscale biventricular models based on subject-specific clinical 12-lead electrocardiogram and magnetic resonance recordings. Essential novel features of the pipeline are the human-based Purkinje network generation method, personalisation considering ECG R wave progression as well as QRS morphology, and translation from reduced-order Eikonal models to equivalent biophysically-detailed monodomain ones. We demonstrate ECG simulations in line with clinical data with clinical image-based multiscale models with Purkinje in four control subjects and two hypertrophic cardiomyopathy patients (simulated and clinical QRS complexes with Pearson's correlation coefficients > 0.7). Our methods also considered possible differences in the density of Purkinje myocardial junctions in the Eikonal-based inference as regional conduction velocities. These differences translated into regional coupling effects between Purkinje and myocardial models in the monodomain formulation. In summary, we demonstrate a digital twin pipeline enabling simulations yielding clinically-consistent ECGs with clinical CMR image-based biventricular multiscale models, including personalised Purkinje in healthy and cardiac disease conditions.

**Graphical abstract**

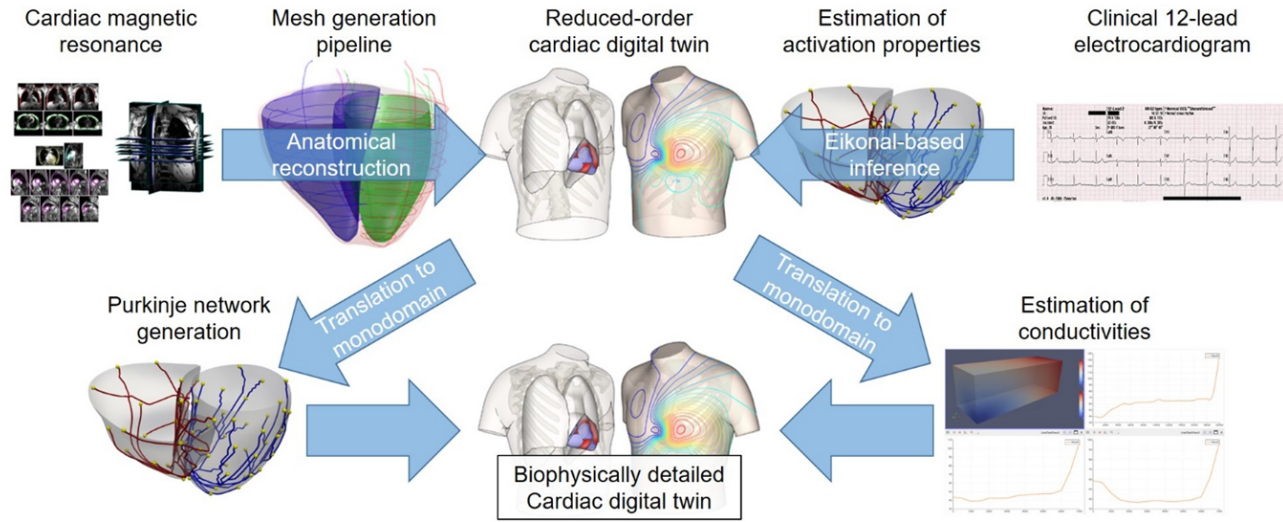

**Highlights**

- Novel pipeline for generating cardiac digital twins, including a Purkinje network from clinical CMR and ECG recordings capable of simulating patient-specific QRS morphological phenotypes.
- Novel human-based Purkinje network generation method to constrain the cardiac digital twins to physiological activation sequences.
- Novel strategy to personalise to ECG R wave progression and QRS morphology.
- Novel strategy to translate reduced-order Eikonal models to equivalent biophysically-detailed monodomain simulations.
- Demonstrated generation of cardiac digital twins for control subjects and disease patients with abnormal QRS complexes.
- The statistical methods presented in this work pave the way for representing electrophysiological uncertainty in cohorts of digital twins for in silico trials.

## 1    Introduction

In silico clinical trials in cardiology is a novel approach to evaluating the safety and efficacy of therapies for cardiovascular diseases (Musuamba et al., 2021; Dasi et al., 2022; Roney et al., 2022). Human-based electrophysiological modelling and simulation can enable identifying risk predictors from therapeutic options more accurately than animal models in some settings (Passini et al., 2017) and provide mechanistic explanations for clinically observed pathological behaviour (Roney et al., 2022; Margara et al., 2022). Similarly to clinical trials, in silico clinical trials entirely rely on the patient cohort representing the variability that would be observed in the human population that each treatment is targeting (Chang et al., 2017; Roney et al., 2022). A challenge in designing these virtual cohorts is the limited information available from clinical data and intersubject variability, requiring novel techniques to reduce and quantify model parameter uncertainty (Musuamba et al., 2021). This uncertainty can be critical in highly inter-subject varying characteristics that have an essential role in the global performance of the heart but are poorly identified through existing clinical tests.

Clinical imaging such as cardiac magnetic resonance imaging (CMR) provides a rich, non-invasive characterisation of cardiac anatomy and structure, widely used in the development of digital twins (Giffard-Roisin, Fovargue, et al., 2017; Lyon, Bueno-Orovio, et al., 2018; Boyle et al., 2019; Roney et al., 2022). Moreover, the 12-lead electrocardiogram (ECG) is the most widely used test clinically, as it supplies a non-invasive characterisation of the cardiac cycle. In particular, the QRS complex of the ECG is the manifestation of the biventricular activation sequence, driven by the conduction system and the Purkinje network. Abnormalities in the Purkinje network are manifested in the QRS complex and are known to promote arrhythmic events and decrease pumping efficiency (Haissaguerre et al., 2016; Ideker et al., 2009). Exploiting the synergies of patient-specific imaging-derived anatomical and ECG-derived electrophysiological information using a digital twinning framework can improve the characterisation and reduce uncertainty on critical properties of the Purkinje network and ventricular activation sequences.

Our study aims to present a novel digital twin generation pipeline to efficiently integrate Purkinje networks in human multiscale ventricular models based on clinical magnetic resonance and 12-lead ECG data. Earlier studies have incorporated Purkinje networks in human cardiac models by combining subject-specific clinical data with physiological knowledge extracted from ex vivo experimental recordings (Kahlmann et al., 2017; Barber et al., 2021; Gillette, Gsell, Bouyssier, et al., 2021). These studies considered reduced-order electrical propagation models based on the Eikonal equation, which can present limitations for in silico clinical trials due to their lack of biophysical detail. Therefore, here we focus on techniques that enable developing and calibrating biophysically-detailed digital twin cohorts, including the Purkinje network, with clinical information available in large datasets, such as the UK Biobank (Sudlow et al., 2015). The pipeline incorporates a novel Purkinje generation strategy coupled with the inference of activation properties to enable variability in the Purkinje network structures in the virtual cohort. It also includes a translation strategy for building personalised biophysically-detailed models from the reduced-order ones made by the inference strategy. We demonstrate the framework using clinical data of four healthy subjects and two patients with hypertrophic cardiomyopathy, a pro-arrhythmic disease, which can lead to anatomical and electrophysiological abnormalities and also affect the Purkinje network (Yokoshiki et al., 2014).

## 2 Materials and Methods

### 2.1 Digital twining framework overview

The digital twinning framework combines a subject's CMR and ECG recordings to estimate a Purkinje network and a set of tissue conductivities that allow replicating the subject's QRS using a monodomain model of human-based electrophysiology. To this end, the framework first reconstructs a 3D geometry of the subject's heart suitable for electrophysiological simulations (Banerjee et al., 2021). Secondly, the framework uses reduced-order (Eikonal) simulations in combination with an inference strategy to estimate a set of earliest activation sites (root nodes for the Purkinje system) and conduction velocities (CV) that reproduce the subject's QRS (Camps et al., 2021). A novelty is the consideration of R wave progression as well as QRS morphology. Finally, the framework translates these root nodes and conduction velocities into a Purkinje network and a set of conductivities suitable for biophysically-detailed (monodomain) simulations of human electrophysiology.

Thus, the digital twinning framework's flow chart (Fig. 1) is composed of the following subprocesses:

1. Automatic subject-specific mesh generation from CMR data (Banerjee et al., 2021).
2. Generation of consistent ventricular coordinates for the subject's CMR-based biventricular geometry (Appendix Section 8.2).
3. Sample a set of candidate endocardial root node locations to be considered by the inference (Section 2.2).
4. Calculate the distance between each candidate root node and the His-bundle in its ventricle while complying with human Purkinje anatomical evidence (Section 2.3).
   *Start of the inference of activation properties (Camps et al., 2021).*
5. Generate the starting population of sets of root nodes and CVs by sampling the activation properties to be inferred and the root nodes with their distances to the His-bundle (Section 2.3).
6. Simulate the local activation time (LAT) maps for each parameter set (Section 2.5) in the population using the Eikonal model (Section 2.2) in combination with the prescribed activation properties (Section 2.5).
7. Calculate the 12-lead QRS complexes from each simulated LAT map (Section 2.2) using the torso-model-derived electrode positions (Section 2.10).
8. Apply the R wave-based normalisation in the calculated 12-lead QRS complexes. (Section 2.4)
9. Evaluate the discrepancies between the simulated and clinical data (Section 2.9).
10. Evaluate the objective criteria.
11. (If criteria are not fulfilled) Update the population of parameter sets (Section 2.5), and continue from 5.
12. (If criteria are fulfilled) Choose the parameter set with the lowest discrepancy.
    *End of the inference of activation properties.*
13. Build a biventricular Purkinje network model complying with the inferred root nodes (Section 2.7).
14. Calibrate the biophysical conductivities to the inferred conduction speeds using human-based cellular models (Section 2.8).
15. Couple the Purkinje network, the biventricular geometry, with the human-based cellular models to produce a cardiac electrophysiological digital twin capable of reproducing the patient's 12-lead QRS complex.

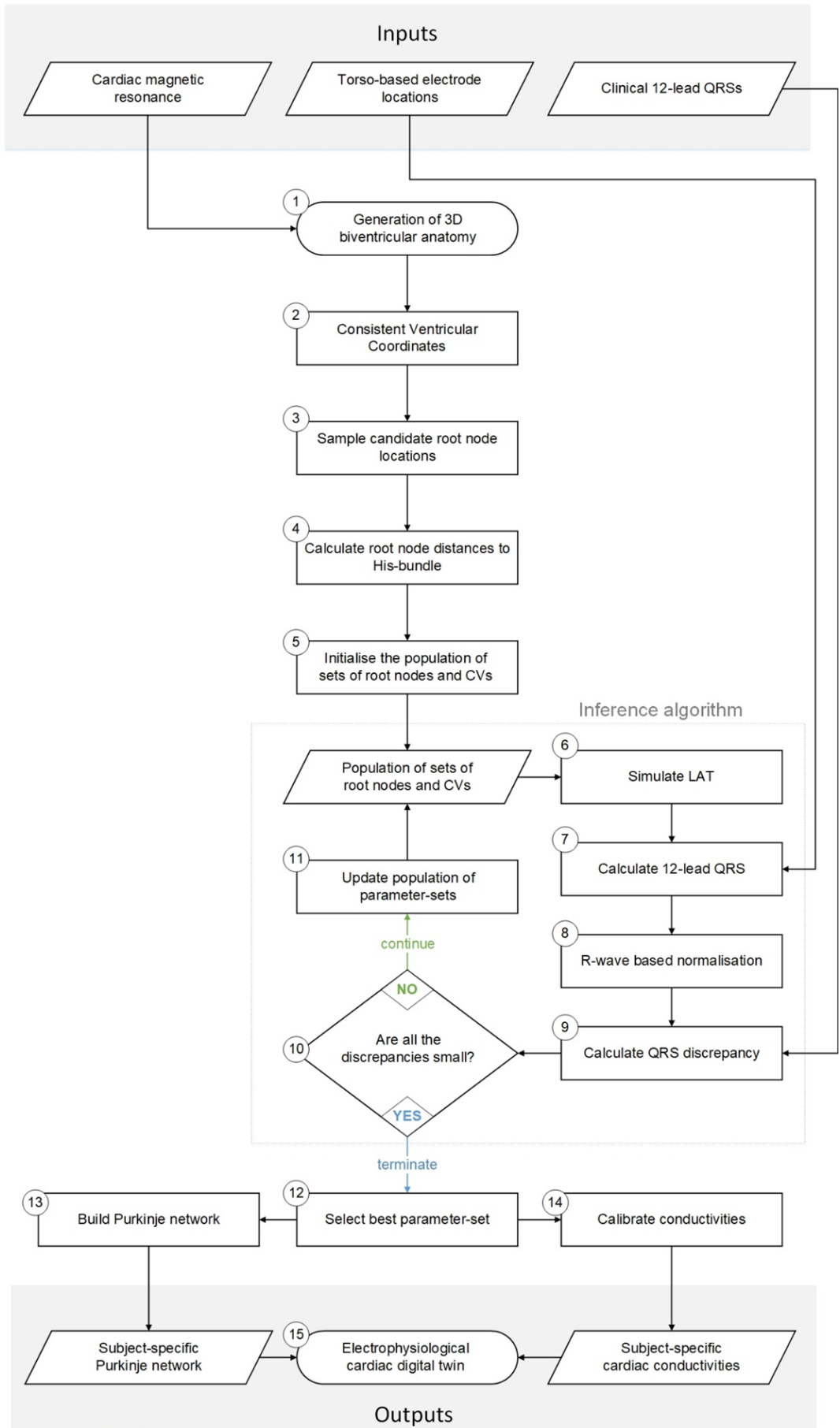

Fig. 1. Digital twin generation framework flow chart. The flow chart outlines the subprocesses comprising the electrophysiological calibration process for a subject given their CMR data, torso-based electrode locations, and clinical ECG. First, the 3D biventricular geometry is reconstructed from the CMR data. Secondly, the consistent ventricular coordinates are generated from the biventricular geometry. Next, the candidate root node locations are sampled by the inference method. Then, the distance between each root node and the His-bundle is calculated accordingly to human Purkinje anatomical evidence. After this step, the population of the parameter set necessary for the inference to calculate the most suitable configuration for the conduction speeds and root nodes are initialised. Then, the inference algorithm begins. At each iteration, the Eikonal simulates the local activation times (LAT), and the pseudo-ECG algorithm calculates the simulated 12-lead QRS complexes; these are then normalised concerning the R wave progression in the subject's clinical ECG and then compared to this same clinical ECG. Finally, the parameter values defining the population of models are updated according to the discrepancy between the simulated and clinical ECGs. This process continues until the population discrepancy is smaller than the termination threshold, in which case, a model from the population is chosen for the biophysical calibration process. These models' parameter values are used to build a 3D Purkinje network and calibrate biophysical parameter values. The resulting calibrated biophysically detailed electrophysiological model, reproducing the subject's clinical QRS, is the final product of the calibration framework. The arrows determine the sequence of events in the flow chart. The shapes symbolise parallelograms – data; rectangles – processes; rhomboids – decisions. The grey background is used for the input and output data in the framework.

### 2.2 Eikonal-based modelling and simulation for inferring root nodes and conduction velocities

Eikonal simulations were conducted for the inference of parameters determining ventricular activation sequence using Dijkstra's algorithm (Dijkstra, 1959), as described in Wallman et al. (2012). The biventricular mesh was considered a connected graph of 1.2 mm edge length, with traversal costs based on orthogonal longitudinal, transmural and normal conduction velocities (CVs) with respect to fibre orientation. Fibre orientations were generated using a rule-based method based on the finding from Streeter et al. (1969).

The endocardial layer was assigned fast isotropic CVs. Two CVs values were considered in the fast endocardial layer, namely 'dense' and 'sparse', to account for the effect of spatial variations in the density of Purkinje-ventricular connections (Myerburg et al., 1972). 'Dense' and 'sparse' endocardial regions were defined using the apex-to-base (ab) and rotation-angle (rt) consistent ventricular coordinates. The 'dense' endocardial region covered the apical areas (ab < 0.4) in the LV and RV, plus the free-wall (0.2 < rt < 0.5), while the rest of the endocardium was considered 'sparse' (Durrer et al., 1970; Myerburg et al., 1972).

The activation sequence was initiated based on root node locations and their activation times. These were determined by first generating branches that connected the root nodes to the His-bundle while following a set of rules extracted from data in the literature. Then, the distances between the His-bundle and each root node through their connecting branches were used in conjunction with the CV through the Purkinje network to calculate activation times at root nodes.

Two possible Purkinje fibres CVs were considered when determining the activation times at the candidate root nodes: 200 cm/s and 400 cm/s, reported as lower and upper bounds of physiological human Purkinje CV (Durrer et al., 1970; Myerburg et al., 1978; Rosen et al., 1981; Joyner & Overholt, 1985; Cragun et al., 1997; Maguy et al., 2009). Additionally, the case of simultaneous activation at the candidate root nodes was also evaluated, as performed by Camps et al. (2021).

The 12-lead ECG-QRS complexes were computed using the pseudo-ECG algorithm (Gima & Rudy, 2002). The QRS complex calculation was obtained by combining the activation time map with a step function to emulate the action potential upstroke.

### 2.3 Purkinje-based Root Node Activation Times

The Purkinje-myocardial junctions in humans are mainly located sub-endocardial (Garcia-Bustos et al., 2019; Vigmond & Stuyvers, 2016), and thus we considered endocardial locations for the root nodes. The algorithm for defining the root node activation times (Camps et al., 2022) is hereafter described. Endocardial surfaces were sampled to obtain equidistant root nodes (2.5/1.5 cm apart, RV/LV) in both cavities. Secondly, Cobiveco ventricular coordinates were considered as transventricular (*tv*) [0, 1] (binary variable), apex-to-base (*ab*) [0 – 1] and rotation-angle (*rt*) [0 – 1] coordinates (Fig. A1) as in (Schuler et al., 2020) (Appendix Section 8.2). Thirdly, the His-bundle in each ventricle was defined at the mid-septal-basal point [1 *ab*, 0.85 *rt*] and the apex at the mid-septal-apical point [0 *ab*, 0.85 *rt*]. Fourthly, the right and left bundle branches were introduced by connecting each bundle branch to its respective apex. This structure serves as the baseline for our 'candidate' Purkinje network to compute root node activation times. Next, four endocardial regions were defined based on how their candidate root nodes connect to the bundle:

- The candidate root nodes in the RV's (1 *tv*) free-wall (0.2 < *rt* < 0.5) and apical (*ab* < 0.2) areas connect to the point in the non-basal bundle (*ab* < 0.8) that shares the most similar *ab* value to them. This creates a rib-caged pattern throughout the RV.
- The apical (1 *tv* & *ab* < 0.2 | 0 *tv* & *ab* < 0.4) and septal (0.7 < *rt*) candidate root nodes in both ventricles connect to their closest apical (1 *tv* & *ab* < 0.2 | 0 *tv* & *ab* < 0.4) point in their bundle. Note that the RV and LV define' apical' regions differently. This rule also applies to the RV's paraseptal (1 *tv*, *rt* < 0.2 | 0.5 < *rt* < 0.7) candidate root nodes.
- The LV's (0 *tv*) paraseptal candidate root nodes (0 *tv*, *ab* > 0.4 & (*rt* < 0.2 | 0.5 < *rt* < 0.7)) connect to their mid-paraseptal-apical point ([0.4, 0.1] or [0.4, 0.6]) and from there to the closest point in the His-bundle (ab < 0.4).
- The non-apical candidate root nodes in the LV's free-wall (0 *tv* & *ab* < 0.4 & 0.2 < *rt* < 0.5) connected to the mid-free-wall-apical point (0 *tv*, 0.4 *ab*, 0.35 *rt*), then to their closest point in the apical (0.4 < *ab*) bundle.

With these routing rules, Dijkstra's algorithm was run on the endocardial surface to generate the paths between each candidate root node and His-bundle. This allows the calculation of the distances considered when activating the candidate root nodes. Finally, activation times at the candidate root nodes were quantified as their distance to the His-bundle divided by Purkinje fibres CV.

### 2.4 Strategies for ECG Comparison

Simulated and clinical ECG comparison was conducted using an improved version of the dynamic time warping algorithm (DTW) (Camps et al., 2021). The DTW algorithm effectively calculates differences between misaligned QRS complexes, which have been normalised using the information from the subject's clinical ECG recording. We extend it by including an R wave-based normalisation in addition to the QRS complexes morphology. The R wave progression of the 12-lead ECG is a well-established clinical biomarker for diagnosis. Furthermore, the R wave progression provides information on the dominant directions of the activation wavefront over time; thus, matching clinical R wave progression can give valuable information for developing cardiac digital twins.

For reporting our results, we adopted the mean ± standard deviation of Pearson's correlation coefficient (Bear et al., 2018; Schaufelberger et al., 2019; Serinagaoglu Dogrusoz et al., 2019) as a measure of disagreement between our inference predictions and the 'target data' (i.e. clinical ECG).

## 2.5 Bayesian Inference and parameter space

The inference method presented here extends Camps et al. (2021, 2022) to account for clinical ECG recordings and consider root nodes with activation time values given by a Purkinje-based model. Briefly, the inference method implements a sequential Monte Carlo approximate Bayesian computation algorithm (SMC-ABC) (Drovandi & Pettitt, 2011) to sample the parameters in the Eikonal model. It uses DTW-based discrepancy to compare the simulated and clinical ECG recordings iteratively, extended to consider R-progression across precordial leads. Finally, we calibrated the hyper-parameters (e.g., weights in the discrepancy metric, the maximum number of iterations, etc.) used by the inference method using subject control-2 with Purkinje speed of 200 cm/s as a reference to enable matching clinical QRS signals. The choice of subject control-2 for hyperparameter calibration was because they displayed the median myocardial and torso volumes across the control subjects considered in this study. These hyperparameter values were then used in all subsequent inferences for the remaining five subjects and the other configurations for subject control-2.

The inference process was conducted under three different assumptions of the ventricular conduction system: Purkinje network with a CV of 200 cm/s, Purkinje network with a CV of 400 cm/s, and simultaneous activation at the root nodes.

The method inferred the value of three continuous parameters: endocardial 'dense' CV, endocardial 'sparse' CV, and transmural CV, as well as the discrete parameters representing the active root node locations. On the other hand, the fibre and normal CVs were considered constant, 65 cm/s and 48 cm/s, respectively, as in Taggart et al. (2000), given their negligible effects on QRS simulations without conduction abnormalities (Camps et al., 2021).

The 'sparse' endocardial CV was allowed to adopt any value within the range [70 – 150 cm/s], the 'dense' CV was allowed to adopt values within [100 – 190 cm/s] (Durrer et al., 1970; Myerburg et al., 1972, 1978), and the transmural CV was allowed within the range [25 – 60 cm/s] (Caldwell et al., 2009; Durrer et al., 1970; Taggart et al., 2000). The number of candidate root nodes changed concerning the endocardial surface area in each geometry. The possible combinations of root nodes ranged from $1.2 \times 10^{10}$ to $8.7 \times 10^{11}$ for the smallest (Control-1) and largest (Control-4) biventricular geometries considered in this study.

The inference method considers a population of 512 parameter sets to allow for identifying different solutions to the inverse problem. Moreover, the inference was repeated three times per configuration to ensure the reproducibility of the results.

## 2.6 Monodomain-based modelling and simulation, including Purkinje network

Monodomain simulations were computed using the GPU-based solver MonoAlg3D (Sachetto Oliveira et al., 2018). The solver implements the finite volume method to solve the monodomain equation on the hexahedral biventricular meshes of the heart. Monodomain simulations were conducted using biophysically-detailed models for human ventricular (Tomek et al., 2019) and Purkinje (Trovato et al., 2020) membrane kinetics. As in the Eikonal simulations, the monodomain ones also considered the rule-based method based on the finding from Streeter et al. (1969) for defining the orientation of the fibres. Based on a convergence analysis using the benchmark mesh from Niederer et al. (2011), which is detailed in Appendix Section 8.7, an 0.4 mm edge-length mesh resolution was considered and 0.4 mm thickness for the endocardial layer, following the same fibre orientation model as in the myocardium (Myerburg et al., 1978). Conductivities were adjusted for each subject following the guidelines presented in Appendix Sections 8.3 and 8.4 to produce an agreement with their clinical QRS duration. This definition of the fast endocardial layer differed from the one considered in the Eikonal simulations. Thus, in the monodomain simulations, the fast endocardial layer will increase the conduction velocity globally, and the Purkinje network will be expected to reproduce the differences across the 'dense' and 'sparse' endocardial regions.

Monodomain simulations were conducted for three conduction system configurations:

1. 'Only root nodes' considered only root nodes and no Purkinje network, similar to the Eikonal simulations employed by the inference process.
2. The 'Minimal Purkinje network' considers only the necessary Purkinje branches to connect to the root nodes.
3. The 'Full Purkinje network' includes further branches and Purkinje-myocardial junctions (PMJs).

In the 'Only root nodes' configuration, the activation sequence was initiated by electrical stimulation of endocardial root nodes. Each root node was modelled as a sphere in the endocardium (*duration* = 4ms and radius = 2mm) with current injected $I_{amp}$ = 53 pA/pF. With the Purkinje network, the stimulus was injected as a single pulse at the proximal $N_{cells}$ of the His-bundle with $I_{amp}$ = 40 pA/pF, *duration* = 4ms and $N_{cells}$ = 25.

In the 'Minimal Purkinje network' and 'Full Purkinje network' configurations, the Purkinje network had an edge-length resolution of 0.1mm (i.e., one hexahedral element).

In the monodomain equation, the surface-to-volume ratio was $\beta$ = 0.14 µm$^{-1}$, and tissue capacitance was $C_m$ = 100 pF/µm$^2$. A total simulation time of $t_{max}$ = 200 ms was used, and a time discretisation $\Delta t$=0.01 ms to solve the diffusion part of the model. To solve the reaction part associated with the Purkinje and ventricular cellular models, a Rush-Larsen scheme of first order was used (Gomes et al., 2020) with a fixed time step of $\Delta t$=0.01 ms to solve the ordinary differential equations for the state variables.

Similarly to the Eikonal model, the QRS from the monodomain simulations were calculated using the simulated membrane potential values, also using a pseudo-ECG equation (Gima & Rudy, 2002).

### 2.7   Construction of human-based Purkinje network anatomical model

The generation of the minimal Purkinje networks for each subject required: i) endocardial surfaces; ii) inferred root nodes with their activation times; iii) Purkinje CV considered in the inference; and iv) an error threshold (i.e., 1 ms) for the translation of the root node activation times. Next, the same rules previously used to define activation times at candidate root nodes were applied to find a new configuration of Purkinje branches (yielding negligible differences in their activation times) and with no branches intersecting. The resulting minimal network connected the inferred root nodes to their ventricular bundle using geodesic lines on the surface of the endocardium. The minimal network from the RV and the LV were joined by their His-bundle, forming a biventricular Purkinje network that can be simultaneously stimulated at a single point and can propagate retrograde electrical waves across ventricles (Fig. 3.A).

The minimal Purkinje network is the baseline for generating a full Purkinje network that physiologically spans across the endocardium. Firstly, equidistant PMJ candidate locations are sampled (0.5/1 cm apart, dense/sparse) (Fig. 3.B). Secondly, the candidate PMJs are connected to the pre-existing Purkinje network one at a time using the algorithm presented in Arantes Berg et al. (2023) (Appendix 8.5), and the activation time of the PMJ is calculated. Only if the new activation time for the new PMJ differs by less than the error threshold (i.e., 2 ms) from the Eikonal simulated activation time for that location, then the new PMJ and branch are added to the pre-existing Purkinje network. Otherwise, the PMJ and its branch are discarded. This process is repeated for all the candidate PMJs, resulting in an expanded version of the minimal Purkinje network with greater coverage (Fig. 3.C). This strategy allows reproducing the inferred differences in the CV across the pre-defined 'dense' and 'sparse' regions as heterogeneities in the density of PMJs.

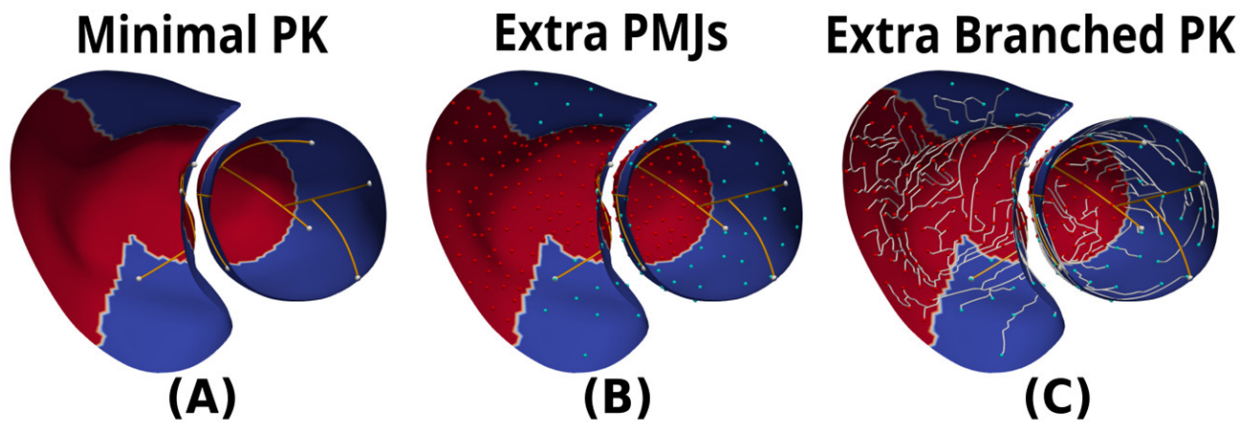

Fig. 3: Illustration of the extra branching procedure. Initially, the minimal Purkinje network (orange) connects the inferred root nodes at their respective LAT. Next, new candidate PMJs are generated over the endocardium's dense (red) and sparse (blue) regions. Finally, the extra branching method is executed considering the minimal Purkinje network as an initial root, generating a dense Purkinje network that connects most of the candidate PMJs at a given error threshold.

The Purkinje-myocardial coupling, which occurs at the PMJs, was modelled by including an additional resistor ($R_{PMJ}$) at the end of terminal branches, and its current $I_{PMJ}$ was injected into tissue elements coupled to the terminal Purkinje elements as described in the following equation:

$$I_{PMJ} = \sum_{i=1}^{N_{PMJ}} \frac{(Vpurk - Vtiss_i)}{R_{PMJ}},$$

where $Vpurk$ is the transmembrane potential of the terminal Purkinje element, $Vtiss_i$ is the transmembrane potential of the tissue element 'I' coupled to the Purkinje element, R$_{PMJ}$ is a fixed resistance, and N$_{PMJ}$ is the maximum number of tissue elements coupled to the Purkinje element, as can be seen in Fig. 2. This additional current is included on the right-hand side of the associated linear system of the ventricular domain only for the elements coupled to the Purkinje system.

The parameters R$_{PMJ}$ and N$_{PMJ}$ were calibrated to reproduce a physiological value for the characteristic 3-25ms anterograde delay that occurs at the PMJ sites (Wiedmann et al., 1996). For all the monodomain simulations, the Purkinje coupling parameter values are set to R$_{PMJ}$=500kΩ and N$_{PMJ}$=60.

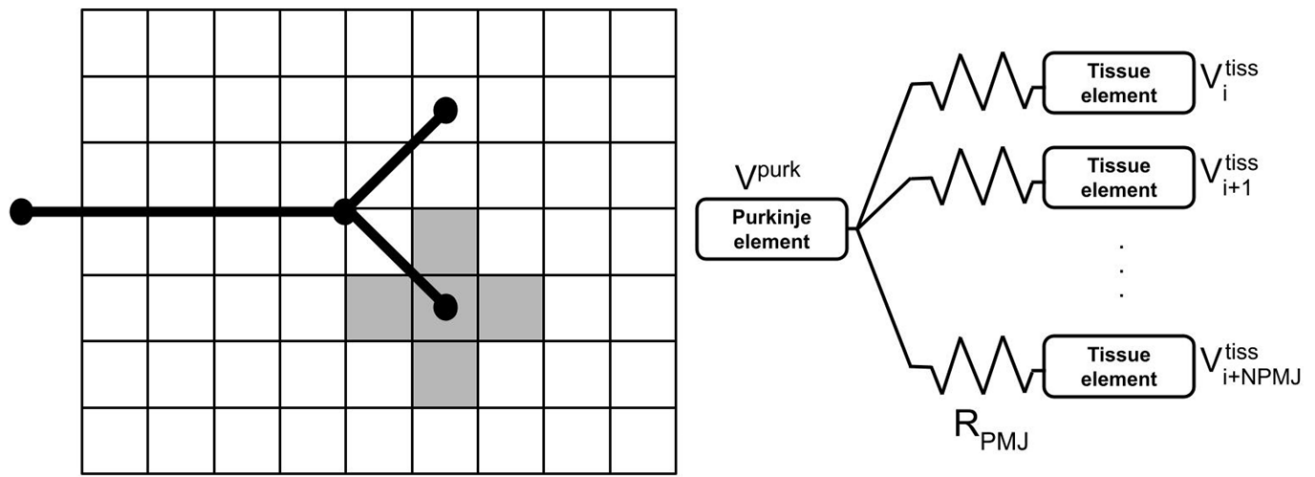

Fig. 2. Illustration of the Purkinje coupling model. The terminal Purkinje elements are linked to the nearest N$_{PMJ}$ tissue elements by a fixed resistance R$_{PMJ}$, and an additional current I$_{PMJ}$ is computed.

**2.8 Translation from Eikonal to Monodomain**

We translated the inferred myocardial and Purkinje network CVs from the Eikonal model by calibrating the conductivities in two monodomain tissue slab setups (Appendix Section 8.3). We calibrated a scaling factor over the myocardial conductivities for the inferred endocardial CVs using a binary search algorithm. This scaling factor was used to ramp up the CV in the fast endocardial layer (one element thick) so that the resulting monodomain simulation matched the QRS duration observed in the subject's clinical 12-lead ECG recording. This procedure was done irrespective of the 'dense' and 'sparse' endocardial regions. The CV differences across these regions were reproduced as changes in PMJ density. The calibrated conductivity values to the inferred conduction speeds using the monodomain tissue slab can be visualised in Table A1 Appendix Section 8.4.

**2.9 Metrics**

To analyse and visualise our root node inference results, we map the errors on bull's eye plots (17 segments per ventricle) (Fig. A1 Appendix Section 8.2). We report local averaged errors between clinical and simulated QRSs as Pearson's correlation coefficient (PCC). We use a normalised (percentage) standard deviation ratio to compare the identifiability across different CVs being inferred. This metric will increase with the differences in the values predicted for each CV across the three repetitions of the inference.

We define the QRS duration of our simulations as the maximum LAT value *minus* the earliest root node activation time in each simulation. The QRS duration for the clinical recordings was considered as the duration of the segmented complexes after the data pre-processing process.

**2.10 Clinical Data and Data Pre-processing**

The digital twinning pipeline is demonstrated using clinical cine CMR and 12-lead ECG recordings from four control subjects and two HCM patients with distinct phenotypes (Lyon, Bueno-Orovio, et al., 2018). The control subjects were selected to display significantly different heart and torso geometries, as in Mincholé et al. (2019), and numbered from one to four in increasing order of their myocardial volume. The HCM subjects had been identified to be representative of the phenotypic HCM subgroups 'group-1A' and 'group-3' (Lyon, Ariga, et al., 2018). Subject 'group-1A' had normal QRS morphology but inverted T-waves and apical and septal hypertrophy caused by repolarisation abnormalities. Subject 'group-3' had abnormally deep and wide S waves and septal hypertrophy, potentially explained by abnormalities in Purkinje-endocardial coupling in Lyon, Bueno-Orovio et al. (2018). These HCM subject subgroups are considered to have high arrhythmic risk. We generated the torso-biventricular 3D geometries from the CMR data (Banerjee et al., 2021; Zacur et al., 2017) (Appendix Section 8.1). All geometries (both control and HCM) were considered to have no regional conduction abnormalities that would affect the activation sequence in the ventricles.

Target clinical ECG recordings were obtained by averaging 20 beats (Lyon, Ariga, et al., 2018) and manually delineating the resulting signals containing just the QRS complex. More precisely, the averaged QRS complexes resulting from the work by Lyon, Ariga, et al. (2018) included additional signals than the QRS and were misaligned across different leads. We corrected the misalignment and trimmed all QRS complexes to begin at the start of the Q wave and end at the end of the S wave while preserving a similar alignment for the R wave and while having the same signal length.

**2.11 Computation and Software**

The inference for the six subjects was conducted with three repetitions, resulting in 18 inferences. Each inference process required about 24 computation hours on a machine with 18x 2nd generation Intel Xeon Scalable Processors.

The inference pipeline was developed in Python/Numpy and can be found at https://github.com/juliacamps/Inference-of-Purkinje-and-ventricular-activation-properties (will be made available after publication). The module to extend the minimal Purkinje network with extra branches, together with the torso-biventricular meshes, are also available under request.

Monodomain simulations were performed with the GPU-based MonoAlg3D software, available at https://github.com/rsachetto/MonoAlg3D_C.

## 3    Results

### 3.1    Personalised Purkinje networks based on ECG and CMR clinical data

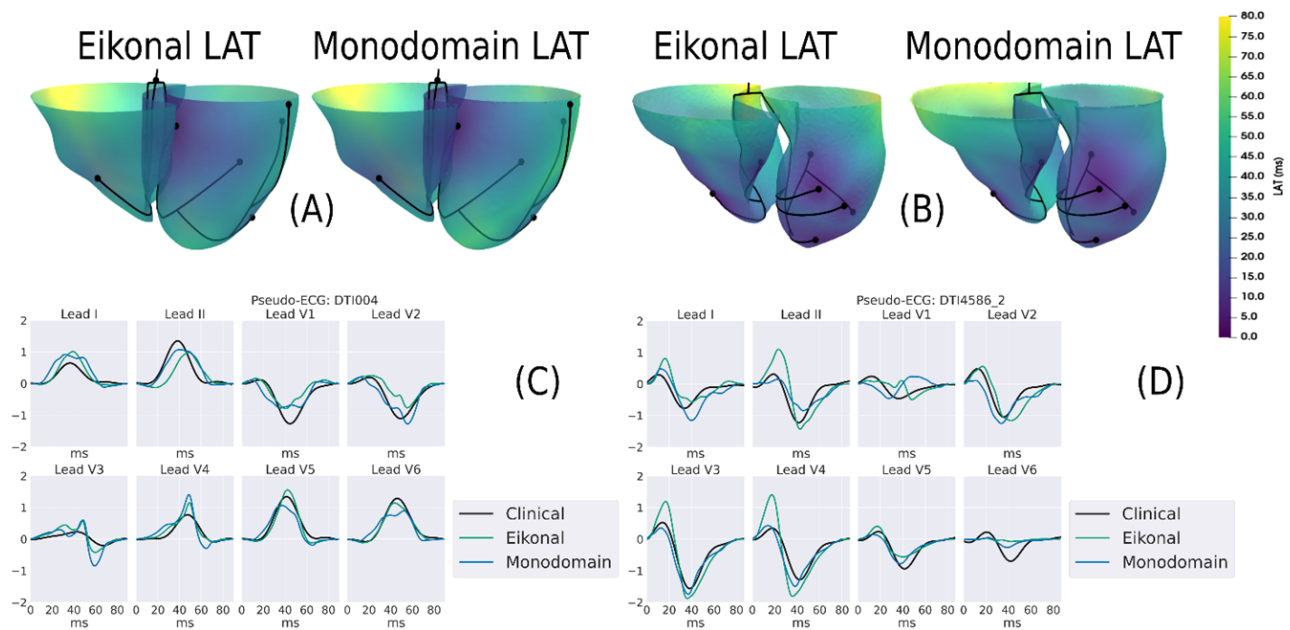

Fig. 4. Personalisation of human-based Purkinje network from clinical QRS and CMR for control and disease subjects. Comparison between simulated endocardial LATs from the inference's Eikonal solution and the monodomain simulations with a minimal Purkinje network (panels A and B for Control-2 and HCM Group 3, respectively). Comparison between simulated QRS complexes and clinical recordings for subjects Control-2 (C) and HCM Group 3 (D).

Fig. 4 and Table 1 demonstrate the ability of the computational pipeline presented above to deliver biophysically detailed biventricular models with Purkinje yielding simulated QRS complexes consistent with clinical ECGs for control subjects and diseased patients using both the Eikonal and monodomain models. Simulated and clinical QRS complexes were very similar, as illustrated in Fig. 4 for Control-2 and HCM Group 3 patients, and quantified in Table 1 for all cases (PCC > 0.8 for most configurations). The best PCC values were achieved when considering either the Purkinje tissue conduction velocity of 400 cm/s or the configuration with simultaneous activation at the root nodes. However, the differences between the average best configurations were smaller than 0.05 PCC in most cases. Moreover, Fig. 4 demonstrates the method's ability to match the clinically observed R wave progression.

| PCC Mean (± std) | Control-1 | Control-2 | Control-3 | Control-4 | HCM Group 1A | HCM Group 3 |
|---|---|---|---|---|---|---|
| PK 200 cm/s | 0.77 (±0.01) | 0.89 (±0.02) | 0.89 (±0.02) | 0.88 (±0.03) | 0.91 (±0.01) | 0.79 (±0.06) |
| PK 400 cm/s | 0.81 (±0.07) | 0.91 (±0.01) | 0.90 (±0.01) | 0.89 (±0.00) | 0.93 (±0.01) | 0.75 (±0.06) |
| Simultaneous RN | 0.85 (±0.03) | 0.90 (±0.00) | 0.93 (±0.02) | 0.91 (±0.02) | 0.92 (±0.01) | 0.87 (±0.01) |

Table 1. Mean ± standard deviation PCC values between clinical and inferred (using Eikonal equation) 12-lead QRS complexes for each subject and three configurations. Row 1: Purkinje network (PK) with a CV of 200 cm/s; Row 2: Purkinje with a CV of 400 cm/s; Row 3: simultaneously activated root nodes (RN). Abbreviations: PK – Purkinje; PCC – Pearson's correlation coefficient; RN – root nodes; HCM – Hypertrophic cardiomyopathy; std – corrected standard deviation.

### 3.2 Incorporation of Purkinje networks while preserving the simulated activation patterns

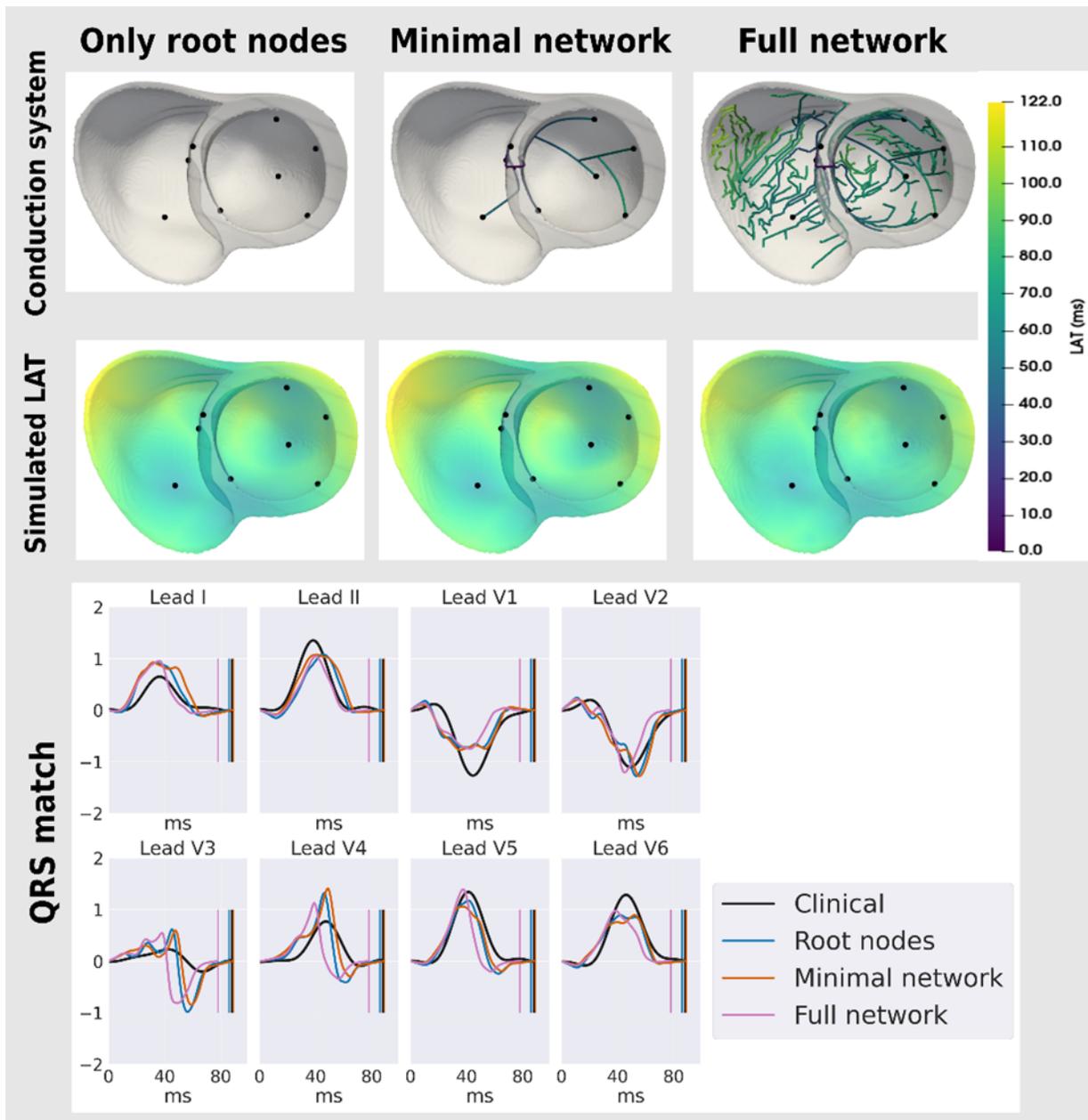

Fig. 5 Effect of the Purkinje network density on the personalised activation sequences in subject control-2 for monodomain simulations. The top row illustrates the modelling strategy for the conduction system: 'Only root nodes' (Root nodes) consider stimulation at the root nodes (black dots) at the times prescribed after the inference; 'Minimal network' considers the necessary Purkinje network to connect the inferred root nodes to the His-bundle and uses this structure to stimulate the ventricles; 'Full Purkinje network' consider a Purkinje network that grows beyond the inferred root nodes. The middle row depicts the simulated LAT for each modelling strategy. Finally, the bottom row illustrates the simulated QRS complexes from 'Only root nodes' (blue), 'Minimal network' (orange), 'Full network' (pink), and the clinical QRS (black).

The patterns in the LAT map were preserved after incorporating the Purkinje stimulation protocol into our personalised monodomain simulations, as well as when extending the Purkinje network to cover the full cavity (Fig. 5). The QRS complexes simulated with the realistic Purkinje networks displayed shorter durations compared to the simulations using the minimal Purkinje network and Only Root nodes setups (Fig. 5). From the results it can be observed that in the monodomain simulations with the Full Purkinje network, the QRS complexes are shorter than the clinical ones due to the electrotonic coupling effects from the additional PMJs, which speed up the already matched QRS width from the Minimal network.

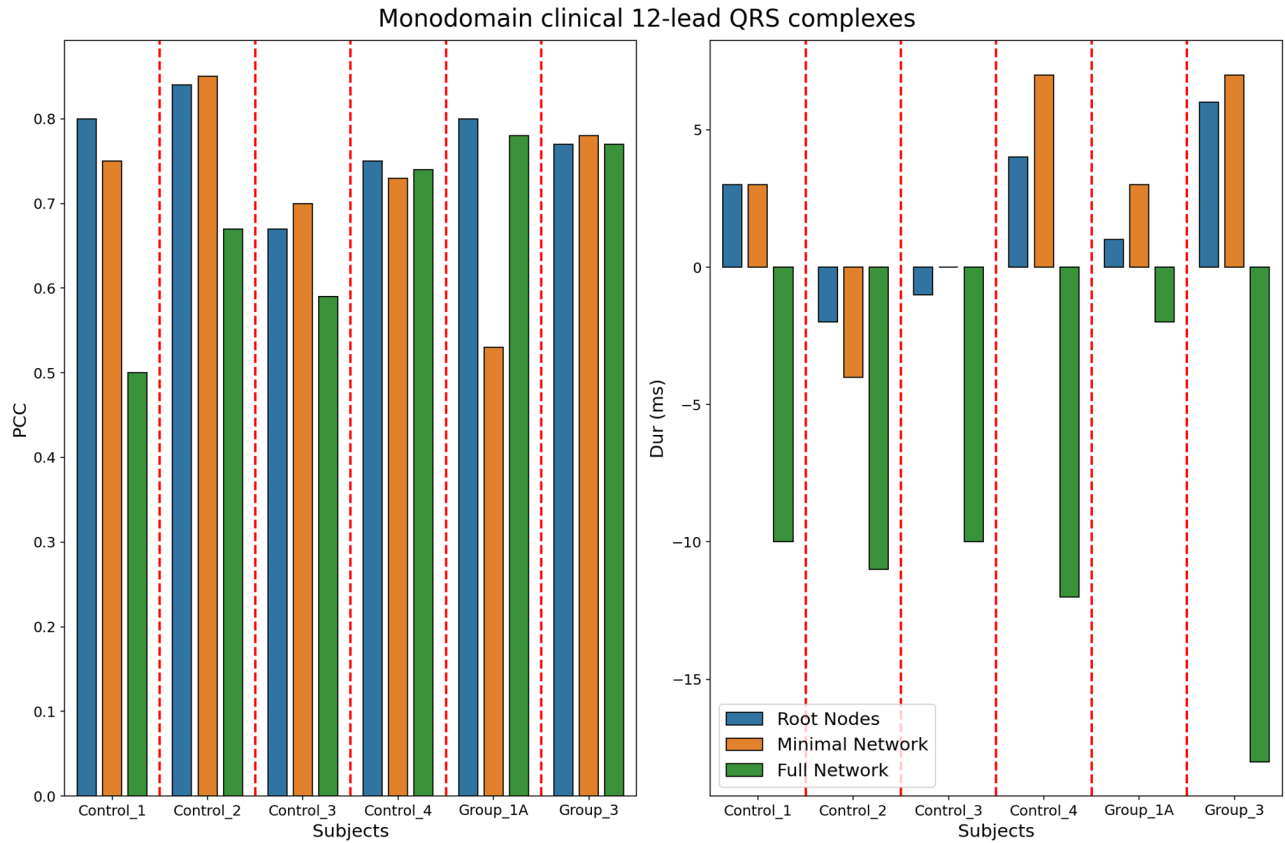

Fig. 6 Pearson correlation coefficients (PCC) and QRS Duration difference (Dur) between monodomain simulations and clinical 12-lead QRS complexes after calibrating the monodomain parameters to the results from the inference with Purkinje CV of 200 cm/s. In the left panel, each colour in the bar plot demonstrates the PCC when modelling the ventricular conduction system to a different extent, while in the right panel, the difference in QRS duration compared to the clinical signal is depicted. 'Only root nodes' (Root Nodes) – baseline calibrated monodomain simulation with stimulation only at the root nodes; 'Minimal network' – incorporates a Purkinje system that only connects from the His-bundle to the root nodes to the baseline setting and stimulates only at the His-bundle; 'Full network' - includes more branches to the Minimal network configuration that covers most of the endocardium.

Simulations using the baseline calibrated monodomain simulation with only root nodes stimulations versus incorporating the minimal Purkinje network achieved a similar PCC match to the clinical 12-lead QRS recordings (Fig. 6). Nevertheless, when expanding the network without further calibration of the monodomain model's parameters, a shortening of the QRS was observed and a subsequent decrease in PCC. The correlation between simulated and clinical QRS complexes using calibrated monodomain simulations was slightly lower than that obtained during the inference of the root nodes from the clinical QRS complex (Table 1).

## 3.3 All conduction velocities were similarly identified for the different configurations of the conduction system

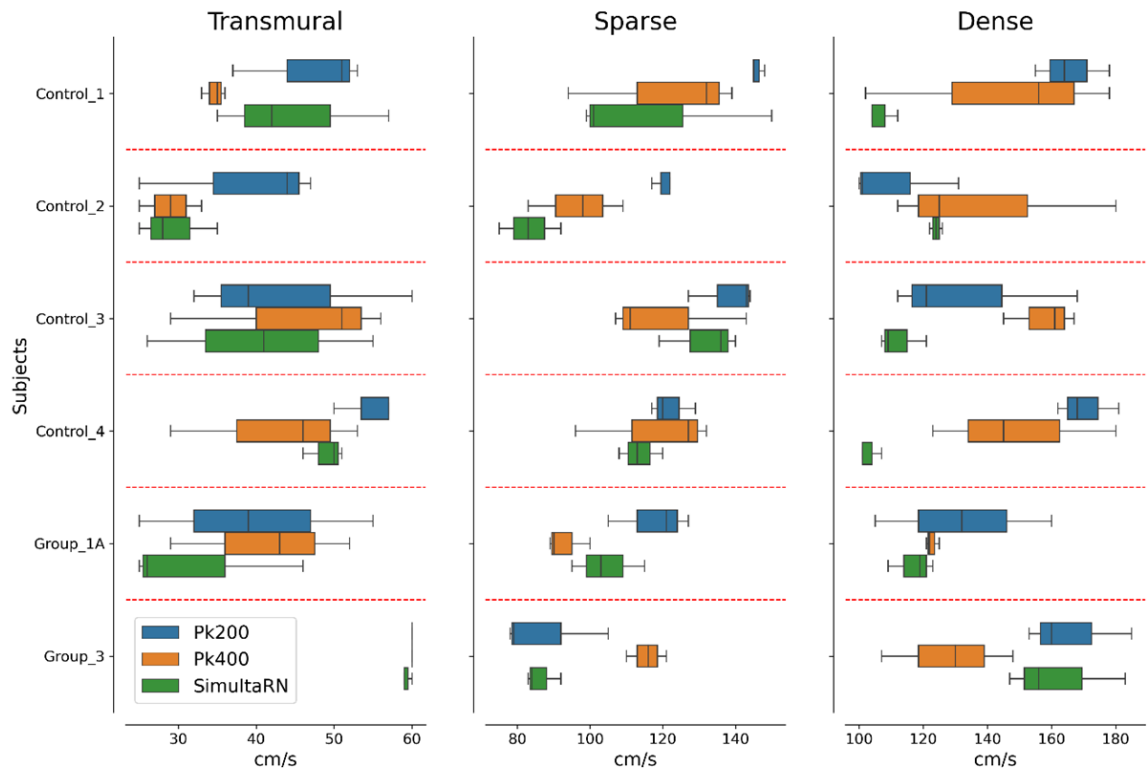

Fig. 7. Boxplot with the inferred CV values for each ventricular region (transmural fibre oriented, sparse and dense fast endocardial layer) considering each subject and their corresponding configuration when using the Eikonal model, where Pk200 is the configuration with a Purkinje CV of 200 cm/s, Pk400 with a Purkinje CV of 400 cm/s and SimultaRN when the root nodes are activated simultaneously.

Fig. 7 shows the CVs were inferred within physiological ranges, [25 – 60 cm/s] for the transmural, [100 – 190 cm/s] dense region and [70 – 150 cm/s] sparse region, for all control and disease subjects when using the Eikonal model (Table A1 Appendix Section 8.4). When the values for the standard deviation are averaged and calculated for each configuration and scenario, we obtain for the 'Transmural' a value of 11.95 cm/s, 12.60 cm/s and 12.95 cm/s for the 'SimultaRN', 'Pk200' and 'Pk400', respectively; for the 'Sparse' scenario the values are 20.55 cm/s, 18.16 cm/s and 21.13 cm/s, respectively; and for the 'Dense' the values are 28.55 cm/s, 25.46 cm/s and 21.59 cm/s.

Overall, from Fig. 7, we didn't identify different trends in the inferred conduction speeds between the control and HCM subject groups. In general, there was no clear relationship between the percentage standard deviation scores among the inference experiments conducted regardless of the CV, cardiac geometry, or stimulation protocol, suggesting similar identifiability capabilities of the CVs across these scenarios. Based on the calculated results from the standard deviation, we can conclude that the 'Transmural' is more identifiable than the 'Sparse,' which is more identifiable than the 'Dense' CVs. This observation is in agreement with the fact that the CVs from the 'Dense' region are the ones that demonstrate higher values and have more variability.

## 3.4 Distribution of root nodes locations is affected by the configuration of inferred CV values

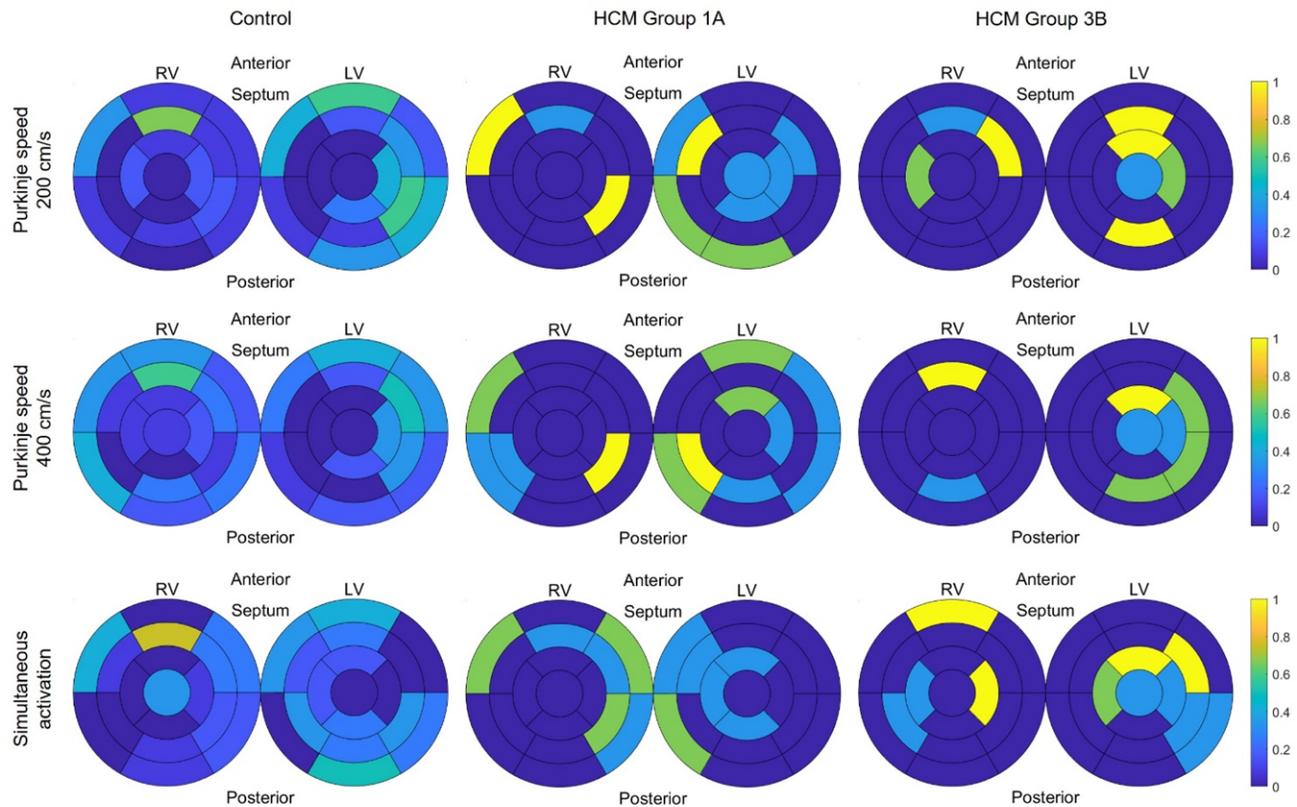

Fig. 8. Comparison of root node locations inferred in control and HCM subjects across all three inference runs per configuration using density maps. The colour map illustrated the percentage of times over the three runs of the algorithm when at least one root node was located in each quadrant/region of the 17x2 endocardial segment chart. The results are combined for all control subjects to be compared with the two representative HCM subjects.

For the control subjects, the inferred root nodes were scattered across the endocardium (Fig. 8). When considering simultaneous root nodes, no root nodes were estimated at the apex or mid to basal anterolateral regions, as can be seen in Fig. 5. The inferred locations of the root nodes were largely affected when we shifted the configuration to activation. In addition, it is essential to notice that in Fig. 8, the results for all control subjects are combined to be compared with the two HCM groups. The average percentage of times when at least one root node was in a quadrant/region is calculated considering the four control subjects.

For the representative HCM subject from group 1A, characterised by having normal QRS morphology, the root nodes shifted from mainly being apical and septal-basal when Purkinje CV was 200 cm/s to anterolateral at the mid and basal when increasing Purkinje CV was 400 cm/s. These root nodes then moved from lateral to septal in both ventricles when further increasing the Purkinje CV to simultaneous activation of the root nodes. The HCM subject from group 3 initially displayed only apical root nodes symmetrically equidistant to the His-bundle. Then, when increasing the Purkinje CV, the inference method estimated more root nodes in the lateral wall, suggesting that the QRS phenotypes from the HCM subject from group 3 required a synchronised activation of apical and lateral regions.

The inferred root nodes from HCM subjects (group 3) were consistently located in the apex region, compared to other subjects (for every Purkinje CV and repetitions of the inference process), and produced deep-wide S-waves in the ECG. This result suggests that a predominantly apical root node location distribution combined with septal hypertrophy can be sufficient to reproduce the phenotypes observed in HCM subjects from group 3.

### 3.5 Simulating a dense Purkinje network speeds up the activation sequence through current injection and coupling effects

Our hypothesis is that during the monodomain simulations, current injection and subsequent electrotonic coupling effects had a major role in QRS shortening simulated from fully grown Purkinje networks compared to their smaller Minimal network counterparts (Fig. 5). The Full network grows using the Minimal network structure as a starting point and the inference LAT result simulation from the Eikonal model as a target to grow additional branches and include new PMJs. Here we test this hypothesis by disentangling the growth of the Full network from the Eikonal simulated LAT. Instead, we use the Minimal network's simulation LAT to generate the Full Purkinje network. Without coupling effects, we would expect at most a 2ms shortening of the QRS complex when extending the Purkinje network given by the LAT error threshold considered in our algorithm.

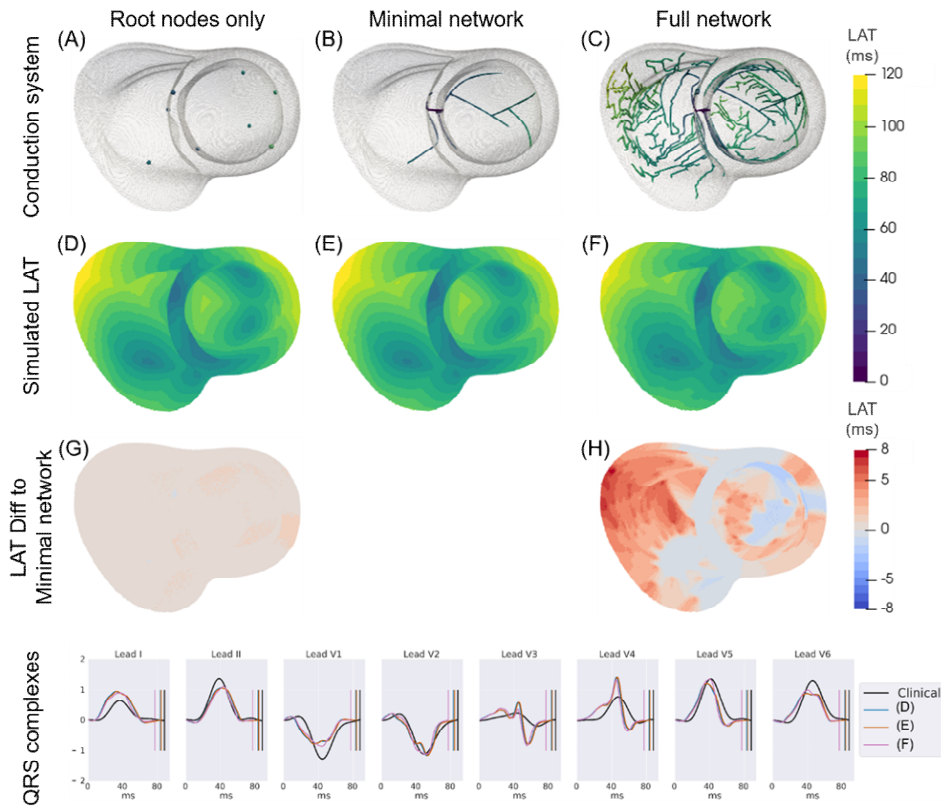

Fig. 9. Comparison of different formulations for the conduction system using the digital twins from control-2. The top row illustrates the modelling strategy for the conduction system: Only root nodes (A), Minimal Purkinje network (Minimal network) (B), and Full Purkinje network (Full network) (C). The second row illustrates the simulated LAT for each configuration considered (D-F). The third row illustrates the difference in simulated LAT maps: (G) between Only root nodes and Minimal network; and (H) between Full and Minimal networks. Thus, positive values indicate that the Minimal network simulation was slower and negative values indicate that it was faster than the conduction systems in the other columns. The bottom row illustrates the simulated QRS complexes from Only nodes only (D, in red), Minimal network (E, in blue), Full network (F, in pink), and the clinical QRS (in black). The coloured vertical lines in the QRS complexes indicate the end of each signal to enhance the visualisation of differences in the simulated signal durations.

The differences in LAT patterns between stimulating the Root Nodes and the Minimal Network simulations were negligible (see Fig. 9). This was also true for their QRS complexes that had a PCC similarity of 0.99 and only a 1ms difference in duration (Root Node simulation had 1ms shorter QRS). This suggests that the stimulus protocol used for the root nodes is similar enough to have PMJs, justifying the translation from one to the other in our digital twin construction pipeline.

The inclusion of additional branches and PMJs to a minimal Purkinje network reduced the time to activate the ventricles by 8.3% (7ms, from 84 to 77ms), which was significantly higher than 2ms (2.4%), i.e., the threshold allowed for additional PMJ (Fig. 9). This suggests that the coupling effects of the additional current delivered from the PMJs are aiding in depolarising neighbouring regions and thus causing a more significant effect than anticipated. On the other hand, we observed a slight slowdown in the activation times neighbouring the original root node locations (Fig. 9 – LAT Diff to Minimal network, third row – third column). This suggests that increasing the network coverage can slightly slows down its CV. These differences resulted in a significant shortening of the QRS duration compared to the minimal network simulation (Fig. 9 – QRS complexes); however, the PCC between these complexes remained at 0.97. All three modelling strategies for the cardiac conduction system obtained similar results to the minimal network configuration (Fig. 9) regarding QRS correlation.

## 4   Discussion

We present a novel digital twin generation pipeline that combines i) an ECG-guided inference method, ii) novel human-based Purkinje representations, and iii) a novel algorithm to extend inference-derived minimal Purkinje networks to realistic, fully grown ones. We also propose a strategy to translate the inference results using reduced-order models into biophysically detailed simulations that match subject-specific clinical recordings. Our pipeline can automatically build biophysically detailed digital twins from clinical CMR and ECG in about 36 hours of computation time. This includes the CMR segmentation and mesh generation process (Banerjee et al., 2021), the Bayesian inference extended from (Camps et al., 2021), the Purkinje system modelling, and the calibration for biophysically-detailed simulations for in silico trials and arrhythmia studies. An important novelty is the ability to match clinical QRS phenotypes beyond QRS duration. Furthermore, our pipeline is suitable for large datasets such as the UK Biobank (Sudlow et al., 2015), generating large cohorts of cardiac digital twins for in-silico trials. We demonstrate the flexibility of our methods and their power in explaining differences in QRS phenotypes across subgroups in the human population through personalisation of the activation sequence. The resulting digital twins matched the morphology and R wave progression of the subject's QRS complexes and enabled personalised mechanistic multiscale simulations. For reproducibility and verification, all codes are openly accessible through this publication.

Our work extends the capabilities of inference pipelines for digital twins in several respects. For example, Giffard-Roisin et al. (2017) estimated the location of just two root nodes in the left ventricle and the CVs from ECG data in a paced heart. Grandits et al. (2020) estimated the activation times of a known set of root nodes. Gillette et al. (2021) inferred activation and repolarisation parameters; however, they simplified the conduction system to have five fixed root nodes with simultaneous activation except for a root node in the free wall (anchoring site of the moderator band), for which they estimated an activation delay constrained between zero and twenty-five milliseconds. Gillette, Gsell, Bouyssier et al. (2021) and Gillette et al. (2022) proposed strategies to extend their previous framework (Gillette, Gsell, Prassl, et al., 2021) to include a realistic Purkinje network without modifying their inference machinery. Improving upon this, our method integrates the Purkinje network design as part of the inference process, using rules derived from human-based experiments in the literature as potential alternatives to initialise the parameter space. Then, it explores the resulting parameter space, which will have an approximate size of the order of $10^{12}$ possible combinations (number of root nodes, their locations and time delays), while also recovering the transmural and two fast endocardial CVs to match the subject's QRS complex in several different physiological ways.

Additional contributions are, firstly, an algorithm for generating Purkinje networks using rules derived from the literature. Secondly, a strategy for expanding minimal Purkinje networks with PMJs covering all regions in the endocardium while preserving the match between clinical and simulated QRS complexes. Secondly, an approach to represent the ventricular fast sub-endocardial layer that mimics effects in the literature on regional heterogeneities of subendocardial conduction velocity through changes in the density of PMJs. This can enable the study of complex pro-arrhythmic wave-propagation behaviours, such as retrograde propagation through the cardiac electrical conducting system. Lastly, a strategy for calibrating biophysically detailed cardiac digital twins from Eikonal-based inference results using GPU-based monodomain simulations of human-based cardiac electrophysiology.

The inference method reproduces all QRS morphologies in diseased and healthy subjects with high fidelity, reporting PCC > 0.85 in at least one configuration of the cardiac conduction system for all subjects. This suggests that our method is flexible enough to represent pathological and healthy subjects without any perceptible biases on cardiac geometry or demographics. Moreover, our updated discrepancy metrics ensure that the resulting QRS complexes have a similar R-progression to the subjects, which improves the clinical relevance of the pipeline compared to our previous work (Camps et al., 2021).

In addition, the inference problem in this work considers a large continuous-discrete mixed parameter space, similar to Camps et al. (2021). However, the major difference in the parameter space for this work is the effect of the non-simultaneous activation of the root nodes. When activated simultaneously, the root nodes have a relative spatial influence on the QRS that depends on the proximity of other root nodes being activated. However, when considering non-simultaneous activations, the root nodes acquire an additional temporal relative influence on the QRS that depends on the activation time of the earliest root node to be chosen. More precisely, in this new formulation, the same root node will affect the QRS at different times depending on its chosen peers. This implies that the first reading in the QRS will come from the earliest root nodes selected in each parameter set. As a less significant difference, this new formulation of the problem allows root nodes not to be used when the activation front reaches the root node before its activation time. Thus, our method can sample 'inactive' root nodes. These differences in the parameter space resulted in a significantly more complex inference problem compared to when considering simultaneous activation at the root nodes in Camps et al. (2021).

We use population parameter sets during the inference to address the non-uniqueness of the solution to this inference problem. Furthermore, to prove the robustness of our results, we repeat the inference three times per configuration and make sure that our predictions are consistent across different runs of the software with different random initialisation of the environment variables. Moreover, the presented method produces a population of solutions to the inference of activation properties (Camps et al., 2021). Although, for demonstration purposes, we only calibrated the biophysically detailed model to the solution with the lowest QRS discrepancy in the configuration, considering a Purkinje network speed of 200 cm/s.

In addition, we were able to calibrate biophysically detailed digital twins for each subject, preserving the high fidelity of the recovered activation sequences that matched the clinical QRS complexes (Fig. 7). The inclusion of a minimal Purkinje network in these cases preserved the match to the clinical data. Nevertheless, growing this network to be fully sized accelerated the propagation of the activation wavefronts through electronic coupling effects (Fig. 5) and subsequently shortened the simulated QRS complexes. Suggesting a new correcting mechanism to counteract this effect is required while growing the Purkinje networks and increasing the number of PMJs in the ventricles.

The non-Purkinje subendocardial tissue can be faster at conducting electrical impulses compared to the rest of the myocardium (Myerburg et al., 1972), especially along the direction of the myocyte fibres in the endocardium (Myerburg et al., 1978) (in canine). The Purkinje network densely covers the endocardial layer in the ventricles and connects to the myocardium with reported densities (in pig hearts) of 0.33 – 0.55 PMJs per mm$^2$ (Garcia-Bustos et al., 2017). On the other hand, our inference method explores root nodes at a density of up to 1 root node per cm$^2$ and can only retrieve up to nine root nodes due to the computational complexity of the inference problem. Moreover, it is known that the density of the PMJs changes across different regions of the endocardium (Garcia-Bustos et al., 2017; Myerburg et al., 1972). Therefore, we defined two density regions based on experiments in Myerburg et al. (1972) since human Purkinje is more similar to canine than pig (De Almeida et al., 2015; Ono et al., 2009). We refer to the region with higher PMJ density as 'dense' and the region with lower PMJ density as 'sparse'.

In our Eikonal simulations, we allow the fast endocardial CV in the 'sparse' and 'dense' regions to be different and isotropic. However, in our monodomain simulations, we define the fast endocardial layer to have a proportionally faster CV than the rest of the myocardium, as suggested by Myerburg et al. (1978). On the other hand, our monodomain simulations account for the differences in CV between these regions by increasing the density of PMJs in the 'dense' region compared to the 'sparse'. The additional PMJs in the 'dense' region will increase the CV by two mechanisms: 1) PMJs can trigger earlier as they are connected through Purkinje tissue that is faster than the fast endocardial layer; thus, a region with additional PMJs will homogeneously activate earlier than without them. 2) The increment in stimulus current delivered from the additional PMJs will boost the depolarisation process in neighbouring tissue through diffusion and locally increase the functional CV during the activation sequence through electronic coupling. The effects of the fast endocardial scaling factor in different conditions on the transmural propagation of the wavefront and on the resulting QRS complexes are then evaluated.

A further experiment was done to evaluate the effect of the fast endocardial scaling factor (Appendix Section 8.6). From these results, we could notice no significant transmural differences in the propagation patterns due to the inclusion of the scaling factor across the different types of conduction systems; thus, the scaling factor had a similar effect to the Eikonal fast endocardial surface layer. In addition, there were minor differences in activation patterns between having only eight stimulation points at the root nodes compared to increasing this number to 1266 points, suggesting that the fast endocardial scaling factor is an appropriate phenomenological surrogate of a denser number of PMJs in our simulations as long as the activation patterns are already well represented. This is also supported by the similar QRS complexes computed from the simulations stimulated at eight and at 1266 locations, which have a PCC similarity of 0.95 (Fig. A2 Appendix Section 8.6).

On the other hand, increasing the number of PMJs by growing the Purkinje network to its full size had a larger effect on accelerating the activation sequence than anticipated (Fig. 5). We expected local increments in the fast endocardial CV because of increases in the density of PMJs. Our strategy was to emulate the CV difference between the 'sparse' and 'dense' regions in the monodomain simulations. However, including the additional PMJs reduced the QRS duration by up to five milliseconds more than the threshold for adding PMJs should allow, demonstrating the extent of the electronic coupling effect on the overall activation sequence. Nevertheless, the morphological traits of the QRS complexes from the minimal network simulation were improved in two subjects after growing the network (Fig. 5); thus, transitioning between these strategies can be done as a post-processing step after the inference process and may require small adjustments of the fast endocardial scaling factor to correct for the CV propagation resulting from the additional current delivered from new PMJs.

The results of the inference evaluation on control and HCM subjects demonstrate the great flexibility of our pipeline in representing varied phenotypes in any subject without tissue structural heterogeneities, such as scars. Interestingly, the inference recovered non-pathological values for the CVs and root node locations for the HCM subject with pathological deep-wide S-waves (group 3), suggesting that abnormal activation properties do not exclusively explain the deep-wide S-wave phenotype describing HCM group 3 but that it is also confounded its septal hypertrophy. We expanded on the findings from Lyon, Bueno-Orovio, et al. (2018) using a targeted digital twin approach that reproduced the clinical data with higher fidelity and explored the parameter space in greater depth. According to our findings, no abnormalities in the conduction system of this subject would be required to reproduce its phenotypical QRS complexes. Nevertheless, given the multiple solution property of this inverse problem, it is uncertain what mechanism may underlie the activation sequence in this group of subjects. Future studies may expand on these findings and compare both mechanistic explanations of the phenotype to clinical and experimental data in the literature and perform the inference in other subjects from the same phenotype subgroup.

Finally, the novel pipeline can be a valuable tool for the calibration process of in-silico trials. For instance, given the patient-specific data, e.g., geometry and ECG, feasible candidates can be selected to replicate the given clinical data, which can be either healthy or under disease conditions.

## 4.1 Limitations

Some of the experimental data used to design the Purkinje strategy and the fibre angle orientations in the heart were based on canine experiments. However, the Purkinje network supports the scarce data available from human experiments. We used a generic configuration for cellular electrophysiology in our monodomain models for this study. In addition, the electrode locations are a source of uncertainty in our inference framework. These were not registered during acquisition and had to be recovered using the subject's torso geometries reconstructed from CMR data. A possible extension of our pipeline could account for the uncertainty in the electrode positions during the inference.

We observed that our inference method favoured the special case of simultaneous activation at the root node locations in terms of PCC match to the clinical QRS and computational cost of the inference process. We suspect the following explanation for this bias: ECG data was pre-processed individually for each lead to align the QRS because the averaging of the beats prevented an alignment of all leads together. This results in a loss of QRS timing differences that might inform activation delays between different spatial sites.

The fidelity of our simulated QRS complexes might have been affected by the uncertainty in electrode locations (Multerer & Pezzuto, 2021) and the distortion errors of the R wave amplitude from using a pseudo-ECG formulation (Nagel et al., 2022). While electrode location is rarely registered in clinical practice, and any formulation of the ECG calculation will introduce some error in the results of the methodology, adopters of our digital twining pipeline can consider known electrode locations and alternative formulations for the calculation of ECG signals without requiring any major changes to the current methods and software. These improvements could result in significantly enhanced matching to clinical QRS complexes.

Pre-defining the activation times at the root nodes constrains the possible structures the Purkinje network may adopt. While the current configuration already leads to a vast parameter space, additional flexibility would enable representing a population of plausible digital twins per subject more likely to include the actual subject's configuration. A possible extension of the current method could include fewer locations for the candidate root nodes while having each root node reached by more than one possible way from the His-bundle. Moreover, the current methodology for the growth of the Purkinje networks operates on a surface and does not allow branch intersection through the Purkinje topology.

## 4.2 Conclusions

We present a novel pipeline for the generation of biophysically detailed cardiac digital twins from CMR and ECG recordings capable of reproducing subject-specific QRS phenotypes in health and disease conditions. Our results demonstrate the need and power of complex digital twinning frameworks in reducing model calibration uncertainty through utilising available clinical data. Our framework can also augment these clinical data to improve risk stratification and mechanistic knowledge of poorly understood disease subgroups through mechanistically integrating multimodal clinical data. Applying this digital twinning framework to larger datasets, such as the UK Biobank, will enable the realisation of the vision for virtual human clinical trials.

## 5 Conflict of Interest

The authors declare that the research was conducted without any commercial or financial relationships that could be construed as a potential conflict of interest.

## 6 Author Contributions

**Julia Camps:** Conceptualization, Methodology, Software, Investigation, Formal analysis, Validation, Visualization, Writing - original draft, Writing - review & editing. **Lucas Arantes Berg:** Conceptualization, Methodology, Software, Investigation, Formal analysis, Validation, Visualization, Writing - original draft, Writing - review & editing. **Zhinuo Jenny Wang:** Conceptualization, Methodology, Software, Formal analysis, Writing - review & editing. **Rafael Sebastian:** Conceptualization, Methodology, Software, Formal analysis, Writing - review & editing. **Leto Luana Riebel:** Methodology, Investigation, Formal analysis, Writing - review & editing. **Ruben Doste:** Software, Formal analysis, Writing - review & editing. **Xin Zhou:** Conceptualization, Formal analysis, Visualization, Writing - review & editing. **Rafael Sachetto:** Software, Formal analysis, Writing - review & editing. **James Coleman:** Formal analysis, Writing - review & editing. **Brodie Lawson:** Methodology, Formal analysis, Writing - review & editing. **Vicente Grau:** Formal analysis, Writing - review & editing. **Kevin Burrage:** Formal analysis, Writing - review & editing. **Alfonso Bueno-Orovio:** Formal analysis, Writing - review & editing. **Rodrigo Weber:** Formal analysis, Writing - review & editing, Supervision. **Blanca Rodriguez:** Conceptualization, Resources, Formal analysis, Writing - review & editing, Supervision.

## 7 Acknowledgements

This work was funded by an Engineering and Physical Sciences Research Council doctoral award, a Wellcome Trust Fellowship in Basic Biomedical Sciences to Blanca Rodriguez (214290/Z/18/Z), the CompBioMed 2 Centre of Excellence in Computational Biomedicine (European Commission Horizon 2020 research and innovation programme, grant agreement No. 823712), the Australian Research Council Centre of Excellence for Mathematical and Statistical Frontiers (CE140100049), an Australian Research Council Discovery Project (DP200102101), by the Queensland University of Technology (QUT) through the Centre for Data Science, a BBSRC PhD scholarship in collaboration with AstraZeneca to Leto L Riebel (BB/V509395/1), and by the Brazilian Government via CAPES, CNPq, FAPEMIG, UFSJ and UFJF, and Generalitat Valenciana Grant AICO/2021/318 (Consolidables 2021) and Grant PID2020-114291RB-I00 funded by MCIN/ 10.13039/501100011033 and by "ERDF A way of making Europe". The computation costs were incurred through a PRACE ICEI project (icp013 and icp019), which provided access to Piz Daint at the Swiss National Supercomputing Centre, Switzerland.

This research was funded in part by the Wellcome Trust [grant number 214290/Z/18/Z]. For the purpose of open access, the author has applied a Creative Commons Attribution (CC BY) public copyright licence to any Author Accepted Manuscript version arising from this submission.

## 8  Appendix

### 8.1  Torso and biventricular geometry reconstructions

The subject's torso-biventricular 3D geometries were generated from cine MR images using the tools demonstrated by Zacur et al. (2017) and Banerjee et al. (2021). This cardiac 3D reconstruction toolkit selects the MR slices at end-diastole and automatically segments them using a UNet machine learning algorithm. These slices are then aligned and corrected to provide smooth transitions and connected to form the epicardial and endocardial surfaces of the left ventricle (LV) and the endocardium of the right ventricle (RV). The RV's epicardium is prescribed using a wall thickness of 4 mm. The resulting biventricular surface meshes are truncated at the base. The torso geometry is reconstructed using the 'localiser' (or 'scout') images acquired at the beginning of the CMR protocol to align the machine with the subject's heart. These scout CMR slices are automatically segmented and aligned. Then, we fit them to a torso statistical shape model to obtain the 3D geometry of each subject. Finally, we used a rule-based strategy to estimate the location of the 12-lead ECG electrodes on the CMR-based torso geometries to enable simulating ECG signals (Zacur et al., 2017).

### 8.2  Biventricular consistent coordinate system coordinate system

To enable the reproducibility and applicability of the work to new subjects, we adopted the consistent biventricular coordinate system Cobiveco (Schuler et al., 2021). It includes symmetric coordinate descriptions of LV and RV and, thus, has endocardial coordinates in the septum of both ventricles. This allows us to define comprehensively the rules for connecting root nodes to their ventricle's His-bundle while highlighting inter-ventricular differences in Purkinje structure (Section 2.3).

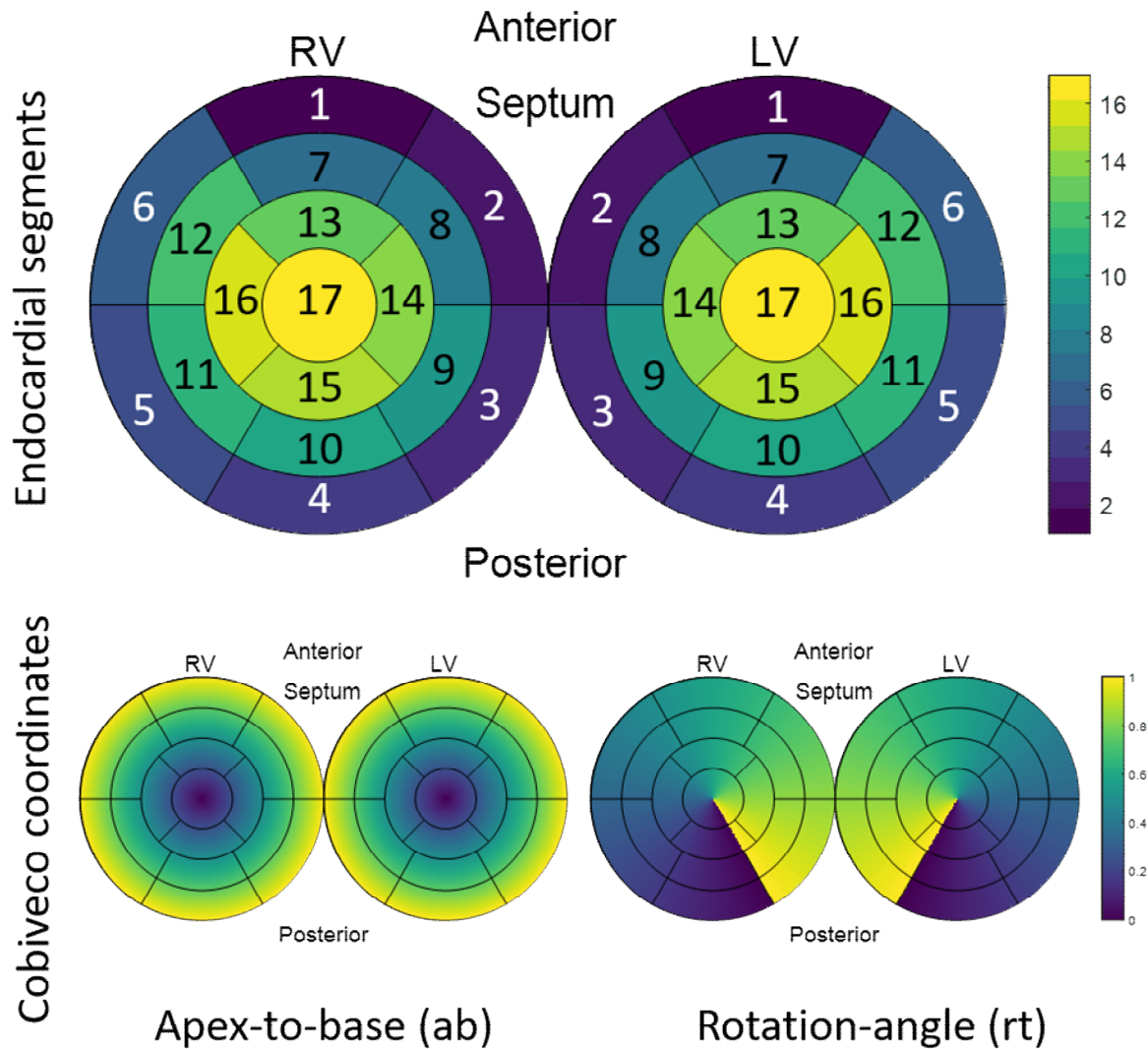

Fig. A1. Symmetric RV and LV endocardial 17 surface segments, similar to the 17-AHA segments (Cerqueira Manuel D. et al., 2002), using Cobiveco's coordinate system (Schuler et al., 2021).

Moreover, in order to increase the interpretability of our inference results, we define a symmetric RV and LV endocardial 17 surface segments (Fig. 2), like the 17-AHA segments (Cerqueira Manuel D. et al., 2002), but just on the endocardial surface, using Cobiveco's coordinate system.

**8.3  Monodomain CV Calibration**

The inferred CVs in the Eikonal formulation cannot be directly translated to the monodomain model. Conversely to the Eikonal CVs model, the values of these conductivities are not independent of each other and depend on the resolution of the mesh in place.

To this end, we defined a 20 x 7 x 7 mm tissue slab aligned with the xyz coordinate system, similar to Niederer et al. (2011). Where the x-axis stands for the y, the fibre in the x-axis, transmural in the y-axis, and normal in the z-axis. The slab was aligned at the origin of the fibre ($f$), transmural ($t$), and normal ($n$) coordinate system (0 $f$, 0 $s$, 0 $n$). This mesh was composed of hexahedral elements at 0.5mm edge-length resolution (same as in the biventricular monodomain simulations). We defined a stimulus region at the corner of the slab (0 *f*, 0 *t*, 0 *n*) with size 1.5 (*f*) x 1.5 (*t*) x 1.5 (*n*) mm. We optimised the conductivity values in all three directions using monodomain simulations with the Tomek et al. (2019) cellular model and applying -50 mV for 2ms to the stimulated region. We computed the CVs along each xyz axis (i.e. fibre, transmural, normal) as the average of first time derivatives of the activation wavefront displacement in the slab. We allowed a discrepancy between the desired CV and the measured value up to 1 cm/s along each direction.

When calibrating the conductivities of the Purkinje network, we used a cable composed of 0.1 mm edged hexahedra of dimensions 100 (*f*) x 0.1 (-) x 0.1 (-) mm, considered 1D for the calibration. This cable implemented human-based Purkinje cell models from Trovato et al. (2020) coupled with a monodomain model. We simulated the activation of this Purkinje cable by stimulating the first 25 elements (2.5 mm long segment) for 2 ms. We calculated the CV using the activation times from the start to end of the 5 mm segment in the middle of the 100 mm long cable. This process is repeated iteratively, changing the tested conductivity value using a scaling factor until the target value for the CV is reached. The process ends when the measured conduction velocity differs from the target by less than 0.1 cm/s. The source code for this process is available in the MonoAlg3D git repository (Section 2.6).

## 8.4 Calibration of monodomain conductivities from Eikonal conduction speeds

|  | Purkinje | | Fast endocardial | | | Fibre | | Transmural | | Normal | |
|---|---|---|---|---|---|---|---|---|---|---|---|
|  | Speed | Conductivity | Sparse cm/s | Dense cm/s | Scaling | Speed | Conductivity | Speed | Conductivity | Speed | Conductivity |
| Control-1 | 200 cm/s | 0.00225 mS | 145 | 164 | 5.67 | 65 cm/s | 0.000310 mS | 53 cm/s | 0.000245 mS | 48 cm/s | 0.000205 mS |
| Control-2 | 200 cm/s | 0.00225 mS | 117 | 101 | 4 | 64 cm/s | 0.000310 mS | 44 cm/s | 0.000185 mS | 47 cm/s | 0.000205 mS |
| Control-3 | 200 cm/s | 0.00225 mS | 127 | 168 | 5.41 | 64 cm/s | 0.000310 mS | 38 cm/s | 0.000155 mS | 47 cm/s | 0.000205 mS |
| Control-4 | 200 cm/s | 0.00225 mS | 129 | 181 | 5.69 | 65 cm/s | 0.000310 mS | 56 cm/s | 0.000270 mS | 48 cm/s | 0.000205 mS |
| Group 1A | 200 cm/s | 0.00225 mS | 121 | 132 | 4.64 | 64 cm/s | 0.000310 mS | 38 cm/s | 0.000155 mS | 47 cm/s | 0.000205 mS |
| Group 3 | 200 cm/s | 0.00225 mS | 78 | 160 | 4.37 | 65 cm/s | 0.000310 mS | 60 cm/s | 0.000295 mS | 48 cm/s | 0.000205 mS |

Table A1. Calibrated conductivity values to the inferred conduction speeds using a monodomain tissue slab (Section 2.8).

## 8.5 Extra Branching Procedure for the Purkinje Networks

To include additional branches to the minimal Purkinje networks, we made use of the method proposed by (Arantes Berg et al., 2023), which can extend an already constructed Purkinje network topology with branches that connect new PMJs using optimisation principles and is represented in Algorithm A1.

---

**Algorithm A1:** Purkinje network extra branching procedure.

---

**Data:** $S$, $x_{prox}$, initial PN.
**Input parameters:** $l_d$, $CF_i$, $CF_a$, $N_i$, $N_a$, $L_{rate}$, $L_{error}$
**Result:** Purkinje network generated within the set of points $S$.

1  $S_i, S_a \leftarrow PreProcessing(S, x_{prox})$ ;
2  $k_{term} \leftarrow RootPlacement(S_i, l_d, x_{prox}, initial\ PN)$ ;
3  **while** ($S_i$ not empty) **do**
4      New intermediate branch $\leftarrow GenerateTerminal(S_i, N_i, CF_i, l_d, k_{term})$ ;
5      Advance to next intermediate point in $S_i$ ;
6      $k_{term} \leftarrow k_{term} + 1$ ;
7      **if** ($k_{term}\ \%\ L_{rate} == 0$) **then**
8          New active branches $\leftarrow AttemptPMJConnection(S_a, N_a, CF_a, L_{error}, l_d, k_{term})$ ;
9      **end**
10 **end**
11 Compute metrics and save network topology to a file ;

---

Algorithm A1. Purkinje network extra branching procedure proposed by (Arantes Berg et al., 2023) with the corresponding subroutines used in this work. To generate an extra branched Purkinje network, the input data is the following: a set of points *S* with distal locations for the terminal branches and containing additional PMJs with their respective LAT to be connected, proximal location of the root branch x$_{prox}$, which indicate the position where the branches start growing and an initial Purkinje network topology represented by the Minimal network previously generated in Section 2.3. The input parameters are: characteristic length of the network, l$_d$, intermediate cost function, CF$_i$, active cost function, CF$_a$, the maximum number of feasible segments to be considered for the intermediate cost function, N$_i$, the maximum number of feasible segments to be considered for the active cost function, N$_a$, connection rate, L$_{rate}$, and error threshold L$_{error}$. The result of the algorithm is a full-branched Purkinje network that tries to connect most of the additional PMJs near their LAT.

In summary, the method loads an initial Purkinje network topology and a set of target PMJs over the endocardium that needs to be connected at a specific LAT. Then it generates new branches to the initial Purkinje network by minimising two cost functions, CF$_i$ and CF$_a$. The first cost function minimises the total length of the network. It is responsible for generating intermediate branches that help the connection of the PMJs. In contrast, the second cost function minimises the LAT error between the Purkinje terminals and the target PMJs and is responsible for generating active PMJ branches. The generation of the branches is intercalated between intermediate and active segments using a connection rate parameter, L$_{rate}$. When the current number of branches in the Purkinje network is divisible by this value, we attempt to connect all the target PMJs using the active cost function, and we only accept the connection if the given error threshold, L$_{error}$, is fulfilled.

In addition, the initial root position of the method, x$_{prox}$, is set to be in the apex node of the left/right bundle branch from the minimal Purkinje network, leading to the generation of branches starting from the apex to the base of the endocardial surface.

The procedure is done for each ventricle separately by loading the topology of the minimal Purkinje network, a subset of points representing the endocardium of the ventricular region, alongside the additional PMJs that will be added to the network with their respective LAT. This subset of points is referred to in the method as S and is later subdivided into the intermediate points set $S_i$ and the active PMJs set $S_a$ by the *PreProcessing* subroutine.

The extra branching procedure initially starts by connecting the active PMJs in the dense region. Next, the resultant topology connects the active PMJs from the sparse region. After generating the two extended Purkinje networks from both ventricles, we link the LV and RV networks together by the His-bundle and consider this full Purkinje network the final topology. Regarding the parameters related to the extra branching procedure, we $l_d$ = 8mm, $N_i$ = 20, $N_a$ = 200, $L_{rate}$ = 25 and $L_{error}$ < 2ms for all the LV and RV networks, where $l_d$ is the characteristic length of the domain, $N_i$ is the maximum number of feasible segments to be considered in the intermediate point cost function, $N_a$ is the maximum number of feasible segments to be considered in the active cost function, $L_{rate}$ is the connection rate constant, and $L_{error}$ is the given LAT error threshold for an active PMJ connection.

In addition, for this study, we allow that some target PMJs cannot connect to the Purkinje network by removing the *PostProcessing* subroutine from the original method. By eliminating this subroutine, we avoid connecting the remaining PMJs left after the main loop with a LAT error not within the LAT error threshold, which could alter the resultant QRS complex to be distinct from the clinical one. It is important to notice that even without this subroutine, we sustain the percentage of successful connections within the LAT error threshold in the main loop to a reasonable value.

## 8.6 The fast endocardial scaling factor in the monodomain simulations preserves transmural isochronic activation bands

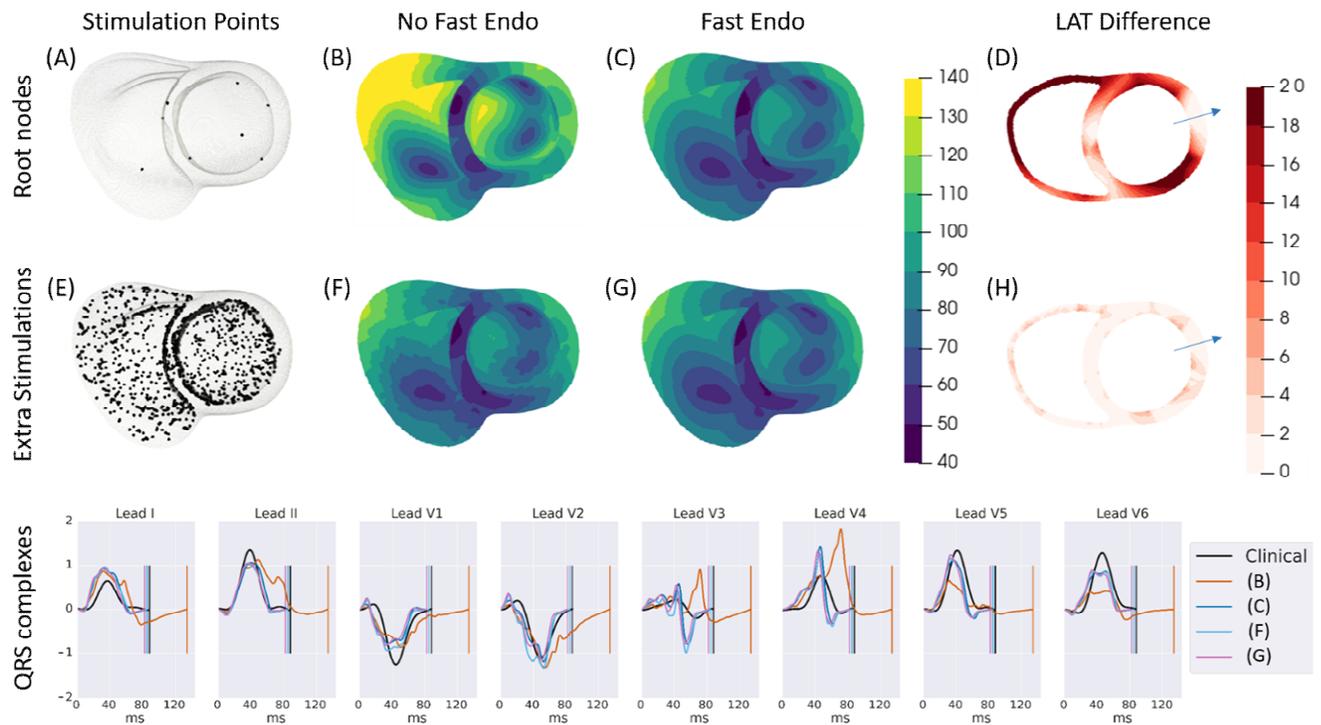

Fig. A2. Effect of including a fast endocardial scaling factor in the monodomain simulations using the anatomy of subject control-2. The first row illustrates the effect of including a fast endocardial layer when only eight root nodes (black dots) are activated. In panel (A), the root nodes are highlighted, while in panels (B) and (C), the LAT maps generated under this configuration without a fast endocardium layer and with the fast endocardium layer are depicted, respectively. In panel (D), the LAT difference, for a cross-section of the mesh, between the two LAT maps shown in panels (B) and (C), namely, the delays in LAT from not having a scaling factor for the fast endocardial layer. Similarly, the second row illustrates the same effect when considering 1266 stimulation points, as depicted in panel (E). This allows observing how the effect of the fast endocardial scaling factor depends on the number and distribution of the stimulation sites. In panels (F) and (G), the LAT maps with and without a fast endocardium layer with 1266 stimulation points, respectively, are shown. In panel (H), the LAT difference, for a cross-section of the mesh, between the LAT maps from panels (F) and (G) are highlighted. Lastly, the third row illustrates the QRS complex simulated from each case. The baseline simulation had only eight root nodes and fast endocardial scaling and corresponded to the calibrated simulation illustrated in Fig. 4. This enables observing these effects concerning the subject's clinical QRS recording.

The inclusion of the fast endocardial scaling enables ramping up the propagation speed in the endocardial layer in our monodomain simulations when considering stimulation only at the eight inferred root node locations (Fig. A2: first row), to match the clinical QRS duration of the subject (Fig. A2: third row). The coupling effects of incorporating this fast endocardial modelling on the transmural propagation patterns are negligible at the location of these root nodes (Fig. A2: third column). The same comparison, when considering 1266 stimulation points (Fig. A2: second row), demonstrates that the differences in activation sequence patterns from the inclusion of the fast endocardial scaling factor are further reduced when increasing the number of stimulation points to be closer to the expected number of PMJs in human ventricles. These findings are supported in terms of the QRS complexes (Fig. A2: third row), where the PCC between having or not the fast endocardial layer is 0.98. Moreover, there were small differences in activation patterns between having only eight stimulation points at the root nodes compared to increasing this number to 1266 points, suggesting that the fast endocardial scaling factor is an appropriate phenomenological surrogate of a denser number of PMJs in our simulations as long as the activation patterns are already well represented. This is also supported by the similar QRS complexes computed from the simulations stimulated at eight and at 1266 locations, which have a PCC similarity of 0.95.

### 8.7 Convergence analysis for mesh resolution

A convergence analysis was done using the benchmark problem established by Niederer et al. (2011) to verify which mesh resolution must be chosen to achieve reasonable results in the monodomain simulations. Here the convergence of conduction velocity in ranges reported by human Taggart et al. (2000) was analysed. To reproduce these conduction velocities, transversely isotropic conductivities $\sigma_l$ = 2.6 mS/cm and $\sigma_t = \sigma_n$ = 1.8 mS/cm were chosen. A tissue slab with the following configuration is set 20 x 7 x 3 mm, with varying values for the mesh resolution (100μm, 150μm, 200μm, 250μm, 300μm, 350μm, 400μm, 450μm and 500μm). A stimulus current is applied in the corner of the tissue 1.5 x 1.5 x 1.5 mm during 2ms and with an $I_{amp}$ = 50 pA/pF.

Regarding the monodomain parameters, $C_m$ = 100pF/cm², β = 0.14 μm$^{-1}$ and the ToR-ORd model were used (Tomek et al., 2019) for the cellular dynamics. The non-adaptive Euler method, in combination with the Rush Larsen method, is used for gating variables with a fixed timestep of Δt=0.01 ms.

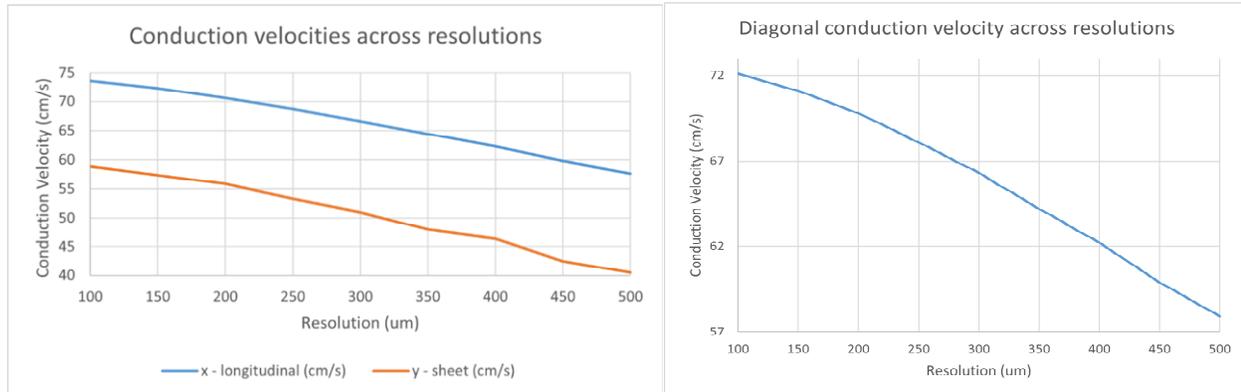

Fig. A2. Conduction velocities are recorded at different mesh resolutions, where in panel (A), the longitudinal (blue) and transverse (yellow) velocities are depicted in cm/s, while in panel (B), the diagonal velocity is highlighted.

Based on the above figure, we conclude that to achieve a discrepancy below 15% in the longitudinal and transverse directions, a mesh resolution of 400μm can be chosen, as can be seen in Fig. A3.

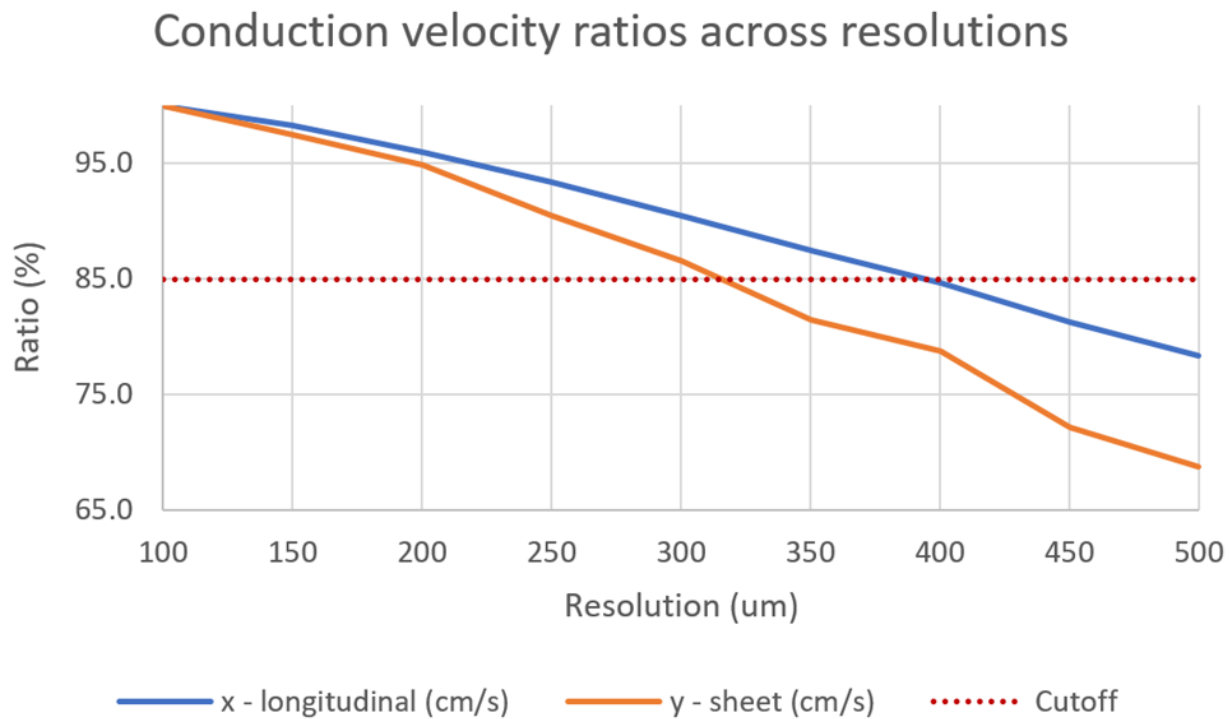

Fig. A3. Conduction velocity ratios across resolutions, where the longitudinal (blue) and transverse (yellow) velocities are depicted in cm/s alongside the 15% discrepancy cutoff.

Running TitleMulterer, M., & Pezzuto, S. (2021). Fast and Accurate Uncertainty Quantification for the ECG with Random Electrodes Location. In D. B. Ennis, L. E. Perotti, & V. Y. Wang (Eds.), *Functional Imaging and Modeling of the Heart* (pp. 561–572). Springer International Publishing. https://doi.org/10.1007/978-3-030-78710-3_54

Musuamba, F. T., Skottheim Rusten, I., Lesage, R., Russo, G., Bursi, R., Emili, L., Wangorsch, G., Manolis, E., Karlsson, K. E., Kulesza, A., Courcelles, E., Boissel, J.-P., Rousseau, C. F., Voisin, E. M., Alessandrello, R., Curado, N., Dall'ara, E., Rodriguez, B., Pappalardo, F., & Geris, L. (2021). Scientific and regulatory evaluation of mechanistic in silico drug and disease models in drug development: Building model credibility. *CPT: Pharmacometrics & Systems Pharmacology*, *10*(8), 804–825. https://doi.org/10.1002/psp4.12669

Myerburg, R. J., Gelband, H., Nilsson, K., Castellanos, A., Morales, A. R., & Bassett, A. L. (1978). The role of canine superficial ventricular muscle fibers in endocardial impulse distribution. *Circulation Research*, *42*(1), 27–35. https://doi.org/10.1161/01.RES.42.1.27

Myerburg, R. J., Nilsson, K., & Gelband, H. (1972). Physiology of Canine Intraventricular Conduction and Endocardial Excitation. *Circulation Research*, *30*(2), 217–243. https://doi.org/10.1161/01.RES.30.2.217

Nagel, C., Espinosa, C. B., Gillette, K., Gsell, M. A. F., Sánchez, J., Plank, G., Dössel, O., & Loewe, A. (2022). *Comparison of propagation models and forward calculation methods on cellular, tissue and organ scale atrial electrophysiology* (arXiv:2203.07776). arXiv. https://doi.org/10.48550/arXiv.2203.07776

Niederer, S. A., Kerfoot, E., Benson, A. P., Bernabeu, M. O., Bernus, O., Bradley, C., Cherry, E. M., Clayton, R., Fenton, F. H., Garny, A., Heidenreich, E., Land, S., Maleckar, M., Pathmanathan,
39